\documentclass{aa}

\usepackage{graphicx}
\usepackage{txfonts}
\usepackage{lipsum}
\usepackage{subcaption}         
\usepackage{lscape}             
\usepackage{placeins}           
\usepackage[normalem]{ulem}
\usepackage{txfonts}
\usepackage{color}
\newcommand{\NB}[2][]{%
  \textbf{\color{red}{#1\if\relax\detokenize{#1}\relax\else\space\fi#2}}%
}

\begin{document}

    \title{The impact of M dwarf stellar convection on exoplanetary atmospheric signatures}
  \subtitle{A 3D radiation-hydrodynamics simulation analysis for high-resolution spectroscopy}


      \author{A. Chiavassa\thanks{Corresponding author:\email{andrea.chiavassa@oca.eu}}
          \inst{1,3}
          \and
          M. Brogi
          \inst{2,3}
          \and
          T. Masseron\inst{4,5}
          \and
          R. Van den Broeck\inst{1}
          \and
          L. Bigot\inst{1}          
          \and
          V. Parmentier\inst{1}    
          }

     \institute{Universit\'e C\^ote d'Azur, Observatoire de la C\^ote d'Azur, CNRS, Lagrange, CS 34229, Nice,  France 
        \and
        Dipartimento di Fisica, Universit\`a degli Studi di Torino,via Pietro Giuria 1, 10125, Torino, Italy
        \and
        INAF, Osservatorio Astrofisico di Torino, Via Osservatorio 20, 10025, Pino Torinese, Italy
        \and
        Instituto de Astrof\'isica de Canarias (IAC), 38205 La Laguna, Tenerife, Spain
        \and
        Departamento de Astrof\'isica, Universidad de La Laguna (ULL), 38206 La Laguna, Tenerife, Spain
        }

   \date{Received 4 July 2026 / Accepted XX November 2026}

 
  \abstract
   {M dwarf stars, as the most abundant and long-lived stellar population in the Galaxy, represent prime targets for the detection and characterization of small, temperate exoplanets; however, their complex activity and convective surfaces pose significant challenges for high-resolution spectroscopic studies aimed at identifying atmospheric habitability signatures.}
   {This work investigates the impact of stellar convection on high-resolution spectroscopy of M dwarfs, quantifies its contamination of exoplanetary atmospheric signals, and develops an integrated framework combining 3D radiative hydrodynamical (RHD) simulations, spectroscopic modeling, and transit analysis to mitigate stellar noise.}
   {We employ 3D RHD simulations of M dwarf convection to compute synthetic spectra using state-of-the-art molecular and atomic line data, and compare them with observations of GJ 3470. We model high-resolution transits of the Neptune-size GJ 3470b, including centre-to-limb variations (CLVs) and the Rossiter--McLaughlin effect (RME), and use principal component analysis (PCA) to de-trend stellar effects.}
   {We show the contribution of different molecular absorption features in the dense and complex spectra of M dwarfs. We calibrate the empirical velocity parameters used in 1D hydrostatic stellar models. Comparison with observations shows that current H$_2$O line lists remain inaccurate, with modest agreement in line strength and position within the $K$-band interval. During the transit of GJ 3470b, the CLVs+RME amplitude is comparable to exoplanet signals. PCA can in principle de-trend these effects, but its behaviour depends on the copresence of telluric and planetary signals and the observational noise budget. While we detect the exoplanet without bias from stellar residuals, this result is limited to a narrow spectral interval and a specific orbital configuration.}
   {This study demonstrates the importance of 3D RHD simulations for modelling convection-induced spectral variability in M dwarfs and provides a framework for quantifying stellar contamination in high-resolution transit spectroscopy, supporting robust atmospheric characterization with  ELT observations.}


   \maketitle

\section{Introduction}


A primary ground-based observational technique to study and characterise the atmospheres of exoplanets is the high-resolution spectroscopy \citep[HRS, e.g., ][]{2025ARA&A..63...83S}. This technique stands out from space-based low-resolution spectroscopy by enabling the unambiguous detection of molecular signals via line-by-line matching, and the observation of Doppler effects arising from planetary motion and possible atmospheric winds. HRS is considered a key ingredient in future exoplanet observations with the Extremely Large Telescope (ELT). It will enable the study of planets around our nearest stellar neighbors in reflected light, allowing measurements of albedo and atmospheric composition at optical and near-infrared wavelengths \citep[e.g.][]{2025ExA....59...29P}. In particular, HRS transmission spectroscopy and reflected light high-contrast spectroscopy may lead to the detection of biomarker signatures in the atmospheres of nearby rocky exoplanets \citep[e.g., ][]{2015A&A...576A..59S,2017A&A...599A..16L}.

M~dwarf stars are the most numerous stellar population in the Galaxy  \citep{2006AJ....132.2360H,2021A&A...650A.201R}. They have masses between 0.08 and 0.65 $M_\odot$, radii 0.10--0.65 $R_\odot$, bolometric luminosities $L_\star$$\sim3\times10^{-4}$--8$\times10^{-2}\,L_\odot$, effective temperatures $T_{\rm eff}$$\sim$2700--4100 K, and surface gravities $\log{g}$$\sim$4.6--5.1 \citep[see][Tables 6, 7]{2015ApJ...804...64M}. Thanks to their low mass and luminosity, they represent prime targets for the study of small and temperate planets \citep{2015ApJ...807...45D,2023A&A...670A..84K}. For this reason, the study of exoplanets orbiting M~dwarf represents the most promising frontier of contemporary astrophysics, with the ultimate goal of identifying potentially habitable worlds with several instruments of the ELT, in temporal order, METIS \citep{2021Msngr.182...22B}, HARMONI \citep{2024MNRAS.528.3509V}, ANDES \citep{2025ExA....59...29P}, and PCS \citep{2021Msngr.182...38K}. M dwarf stars with masses between 0.08 and 0.4 M$_{\odot}$ are fully convective and can generate strong magnetic fields \citep{2008ApJ...676.1262B,2015ApJ...813L..31Y}. Their low effective temperatures result in a plethora of molecular line transitions throughout the optical spectrum \citep[e.g. TiO, VO, ZrO, CO, H$_2$O, MgH, OH, CaH, FeH][]{2018A&A...610A..19R,2021A&A...656A.162M} and atomic line transitions \citep[e.g.][]{2018A&A...620A.180R} for which non-LTE effects on particular potassium strong lines lead to non-negligible metallicity corrections \citep{2021A&A...649A.103O}. M dwarf stars exhibit conspicuous and abundant evidence of surface activity: flares, photometric rotational variability, enhanced chromospheric and coronal emission in X-rays, UV, and radio that are believed to be correlated with the presence of intense magnetic field \citep[for a detailed review see ][]{2021A&ARv..29....1K}. In addition to stellar activity, the surface of these fully convective stars is covered with complex, stochastic patterns associated with convective heat transport, the stellar granulation. Convection manifests in the surface layers as a particular pattern of downflowing cooler plasma and bright areas where hot plasma rises \citep{2009LRSP....6....2N}. 
The direct consequence of the non-stationary stellar flux is a non-stationary spectrum, either a Doppler shift or a distortion of the line profile during planetary transits, producing a non-negligible source of noise that can alter or even prevent detection, especially when the same spectral lines are present both in the stellar and in the planet spectrum \citep{2019A&A...631A.100C}. In this context, and thanks to its capability to separate signals with distinct Doppler signatures, HRS is especially effective for identifying potential sources of spurious signals that can complicate the interpretation of exoplanet spectra. It is thus crucial to have an accurate and physical representation of time-dependent stellar disk background phenomena using three-dimensional (3D) numerical radiative hydrodynamical (RHD) simulations of M dwarf stellar convection. Pioneering works on M dwarfs have shown the challenges in modelling these objects with respect to  main sequence stars and have highlighted the presence of self-excited gravity waves as an essential mixing process in their atmospheres \citep{2002A&A...395...99L,2010A&A...513A..19F}. More recently, 3D simulations covering the stellar parameters of M dwarf stars have been developed \citep[e.g. ][]{2009A&A...508.1429W,2024A&A...688A.212R,2025ApJ...978L..13S}. Theoretical efforts for obtaining detailed line formation physics, using state-of-the-art time-dependent hydrodynamics simulations, are crucial in determining spectroscopic-based parameters on these stars as well as their impact on atmospheric signatures of planets. 


In this work, we report a complete analysis of the synthetic spectra computed from new 3D RHD simulations using state-of-the-art molecular and atomic data. We estimate the impact of convection-related signatures on stellar spectra and report a spectral analysis of molecular data from M-dwarf. Moreover, we present the first integrated approach that combines stellar convection simulations, spectroscopic modeling, and transit analysis, with the aim to master and reduce the stellar contamination in exoplanet atmosphere observations.


\section{Stellar surface convection}

\subsection{3D radiation-hydrodynamic simulations}

We used  the \texttt{Stagger} code to calculate the simulation of stellar convection in M dwarf stars. \texttt{Stagger} is a compressible radiation-(magneto)hydrodynamic
code that solves the time-dependent hydrodynamic
equations for the conservation of mass, momentum, and energy, together with the radiative transfer equation, assuming local thermodynamic equilibrium (LTE). The code was originally developed by \cite{1995stagger} and further improvements were made in \cite{2013A&A...557A..26M} and \cite{2018MNRAS.475.3369C}. A recent comprehensive description of the code can be found by \cite{2024ApJ...970...24S}.

The radiation transfer is solved along a total of nine rays, combining one vertical ray, two inclined polar angles, and four azimuthal angles, for a total of eighteen directions \citep{2003ASPC..288..519S}.
The full computation across all wavelength/frequencies is computationally not possible; thus, \texttt{Stagger} employs the opacity binning method \citep{1982A&A...107....1N,2000ApJ...536..465S,2013A&A...557A..26M,2018MNRAS.475.3369C}: the radiative transfer is calculated for a large number of wavelength points merged into 12 opacity bins. They include continuous absorption opacities and scattering coefficients from \cite{2010A&A...517A..49H} as well as (molecular and atomic) line opacities described in \cite{2008A&A...486..951G}. A solar chemical composition is assumed \citep{2009ARA&A..47..481A}.

\begin{figure}[t!]
   \centering
   \includegraphics[width=\hsize]{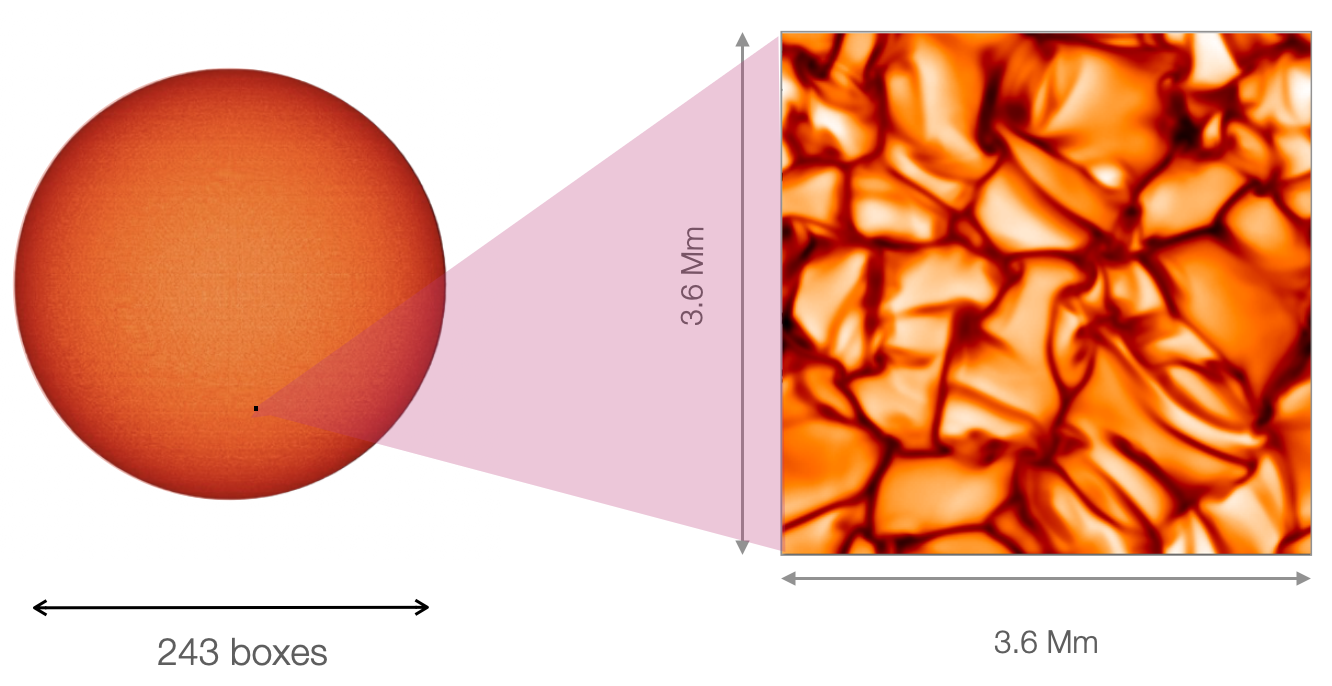}
      \caption{Bolometric intensity map (right) from the M dwarf simulation of Table~\ref{simus}. The sketch on the left shows the approximate size of the numerical box with respect to the stellar radius, assuming a \cite[$R$=0.4 $R_\odot$ for a 3500K M dwarf star,][]{2018MNRAS.481.1083P}.}
         \label{figsimus}
   \end{figure}

As a consequence of the use of \texttt{Stagger} and similar codes, a massive construction of extensive grids of 3D simulations has resulted in models covering a large part of the HR-diagram \citep[e.g., ][]{2009MmSAI..80..711L,2013A&A...559A.104T,2013A&A...558A..49B,2013A&A...557A..26M,2024A&A...688A..52E,2024A&A...688A.212R}. In particular, the  \texttt{Stagger}-grid originally contained about 200 models covering a broad range in stellar parameters but still above the M~dwarfs effective temperature and surface gravity regime \citep{2013A&A...557A..26M}. \cite{2024A&A...688A.212R} updated the grid by including new cool dwarf simulations. In this work, we use a representative M~dwarf simulation (Table~\ref{simus}) from the \texttt{Stagger}-grid to compute synthetic spectra. The simulation size with respect to the stellar radius is shown in Fig.~\ref{figsimus}. The total simulated time is about 2 hours to capture several convective turnovers.

\begin{table*}
\centering
\begin{minipage}[t]{\textwidth}
\caption{The M~dwarf models used in this work.}           
\label{simus}      
\centering                          
\renewcommand{\footnoterule}{} 
\begin{tabular}{l c c c c c c c c c}        
\hline\hline                 
3D/1D & $<T_{\rm{eff}}>$\tablefootmark{a} & [Fe/H]  & $\log{g}$ & $x,y,z$-dimensions & $x,y,z$-resolution   & $\rm{M}_{\star}$ & $\rm{R}_{\star}$ & granule size \tablefootmark{b}  & Number of tiles \tablefootmark{c} \\
& $[\rm{K}]$ & & [cgs]  & [Mm]  & [grid points]   & [$\rm{M}_\odot$] & [$\rm{R}_\odot$] & [$10^{-3}$] & over the diameter\\
\hline
3D & 3500 & 0.0 & 4.5 & 3.60$\times$3.60$\times$1.04 & 120$\times$120$\times$110 & 0.4 & 0.45 & 2.87 & 243 \\
1D & 3500 & 0.0 & 4.5 & ... & ... & ... & ... & ... & ... \\
\hline                          
\end{tabular}
\tablefoot{The 3D RHD simulation comes from \texttt{Stagger}-grid \citep{2024A&A...688A.212R}. The total simulated time is $\approx$2 hours. The 1D hydrostatic plane-parallel MARCS model has been computed with a microturbulence of 1 km\,s$^{-1}$ and comes from https://marcs.astro.uu.se \\
\tablefoottext{a}{Horizontal and temporal average of the emerging effective temperatures}
\tablefoottext{b}{from the intensity map of Fig.~\ref{figsimus} divided by the stellar radius}
\tablefoottext{c}{$N_{\rm{tile}} = \frac{\pi \cdot \rm{R}_\star}{x,y\rm{-dimension}}$ from \cite{2017A&A...597A..94C}}
\tablefoottext{d}{for a 3500\,K M dwarf star \citep{2018MNRAS.481.1083P}}
}
\end{minipage}
\end{table*}

\subsection{Synthetic spectra}\label{sec:spectra}

We used the 3D pure-LTE radiative transfer code \texttt{Optim3D} \citep{2009A&A...506.1351C} to compute synthetic spectra from the snapshots of the RHD simulation reported in Table~\ref{simus}.
The code takes into account the Doppler shifts due to convective motions. The radiative transfer equation is solved monochromatically using pre-tabulated extinction coefficients as a function of temperature, density, and wavelength. The lookup tables were computed for the same chemical compositions as the RHD simulations using the extensive atomic and molecular continuum reported in Table~\ref{molatoms} and the continuum opacities are from \citet{2008A&A...486..951G}. We assumed microturbulence equal to zero since the velocity fields inherent in RHD simulation is expected to self-consistently and adequately account for non-thermal Doppler broadening of spectral lines \citep{2000A&A...359..755A}. More details about \texttt{Optim3D} can be found in \citet{2009A&A...506.1351C,2010A&A...524A..93C}.

\begin{table}
\tiny
\centering
\begin{minipage}[t]{\columnwidth}
\caption{Sources of data used for the post-processing calculations with \texttt{Optim3D} for molecular and atomic line opacities}             
\label{molatoms}      
\centering                          
\renewcommand{\footnoterule}{} 
\begin{tabular}{l l}        
\hline\hline                 
Species & References \\
\hline
Atomic lines & severals\tablefoottext{a} \\
AlH & \cite{Yurchenko2018_AlH} \\
AlO & \cite{2015MNRAS.449.3613P} \\
C$_{2}$ & \cite{Yurchenko2018_C2} \\
CaH & \cite{Yadin2012,2012MNRAS.425...34Y,Shayesteh2013} \\
CH & \cite{Masseron2014} \\
CrH & \cite{Burrows2002} \\
CN & \cite{Sneden2014} \\
FeH\tablefootmark{b}  & \cite{Dulick2003, Hargreaves2010} \\
H$_2$O  Pokatazel & \cite{2018MNRAS.480.2597P} \\
H$_2$O  MARVEL\tablefootmark{c} &  \cite{2012JQSRT.113..929F} \\
H$_2$O  HITRAN &   \cite{2010JQSRT.111.2139R} \\
H$_2$O  Barber &   \cite{2006MNRAS.368.1087B} \\
LaO & \cite{2022ApJ...933...99B,2023ApJ...953..181B} \\
MgH & \cite{Yadin2012,2012MNRAS.425...34Y,Hinkle2013} \\
NH & \cite{2018JQSRT.217...29F} \\
OH & \cite{Brooke2016,Yousefi2018} \\
SiH & \cite{Yurchenko2018_SiH} \\
SiO & \cite{Barton2013} \\
TiO & \cite{McKemmish2019} \\
VO & \cite{McKemmish2016} \\
\hline                          
\end{tabular}
\tablefoot{\tablefoottext{a}{\cite{Smith2021} and \cite{Heiter2021}. A sub-list including only iron lines in the visible is used in Sect.~\ref{micromacrosection}}
\tablefoottext{b}{The FeH line list encompasses the $F\,^4\Delta$--$X\,^4\Delta$ and $E\,^4\Pi$--$A\,^4\Pi$ electronic systems, taken from \citet{Dulick2003} and \citet{Hargreaves2010}, respectively. For the $E$--$A$ system we adopt the empirically adjusted $gf$-values of \citet{Souto2017}; these are confined to the $H$ band, as the list was assembled for the APOGEE survey \citep{Smith2021}.}
\tablefoottext{c}{It is a sub-list of Pokatazel, whose energy levels have been adjusted and potentially improved with this algorithm.} 
}
\end{minipage}
\end{table}

\begin{figure}
   \centering
   \includegraphics[width=1.\hsize]{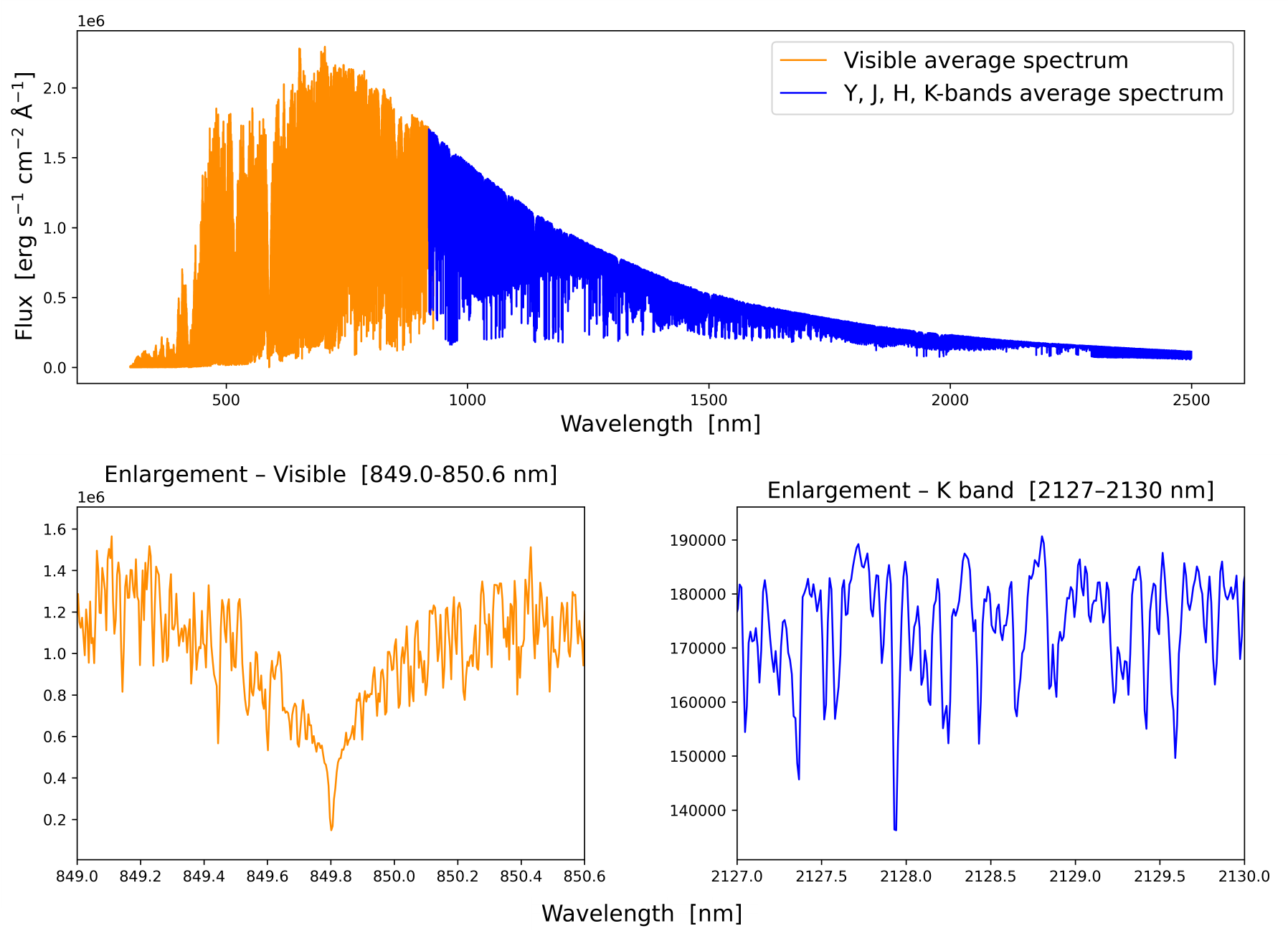}
      \caption{Temporal and spatially averaged synthetic spectrum of the RHD simulation from Table~\ref{simus} across the visible and infrared regions. It includes the molecular and the atomic lists of Table~\ref{molatoms}. The bottom panels show two examples of enlargement for a Ca~{\sc ii} line (left) and the spectral region analyzed in Sect.~\ref{sec:clv+rme_model} (right)}
         \label{figspectrglo}
   \end{figure}

We computed spectra from the 3D RHD simulation (Table~\ref{simus}) with constant resolving power $\lambda/\Delta\lambda$=200,000 from 300 to 2500 nm (Fig.~\ref{figspectrglo}), and covering the visible and infrared part of the stellar spectrum. The radiation transfer is done by casting vertical rays through the computational box for all required wavelengths and using ten box tilting angles $\mu_{10}=\cos(\theta)$= [1.00, 0.90, 0.80, 0.70, 0.50, 0.30, 0.20, 0.10, 0.05, 0.01], where $\theta$ is the angle between the normal to the surface and the line of sight, and four azimuthal rotations $\phi$ = [0$^\circ$, 90$^\circ$, 180$^\circ$, 270$^\circ$] in Sect.~\ref{dwarfsynthetic}. We used instead 43 angles $\mu_{43}$ = [1.00000, 0.996358, 0.992716, 0.989074, 0.985432, 0.981790, 0.978148, 0.967381, 0.956614, 0.945847, 0.935080, 0.924313, 0.913545, 0.896124, 0.878702, 0.861281, 0.843860, 0.826438, 0.809017, 0.785703, 0.762388, 0.739074, 0.715760, 0.692445, 0.669131, 0.640942, 0.612754, 0.584565, 0.556377, 0.528188, 0.500000, 0.468169, 0.436339, 0.404508, 0.372678, 0.340847, 0.309017, 0.274936, 0.240854, 0.206773, 0.172691, 0.138610, 0.104528] in Sect.~\ref{sec:clv+rme_model}. In the case of 10 $\mu$ angles, the number of synthetic spectra is 120 with a step of 60~s for total time of 2 hours of stellar convection time (Table~\ref{simus}). We also computed the temporal variability, and we performed temporal averages using all the snapshots.

\section{M~dwarf synthetic spectra}\label{dwarfsynthetic}

\begin{figure*}
   \centering
     \includegraphics[width=0.85\hsize]{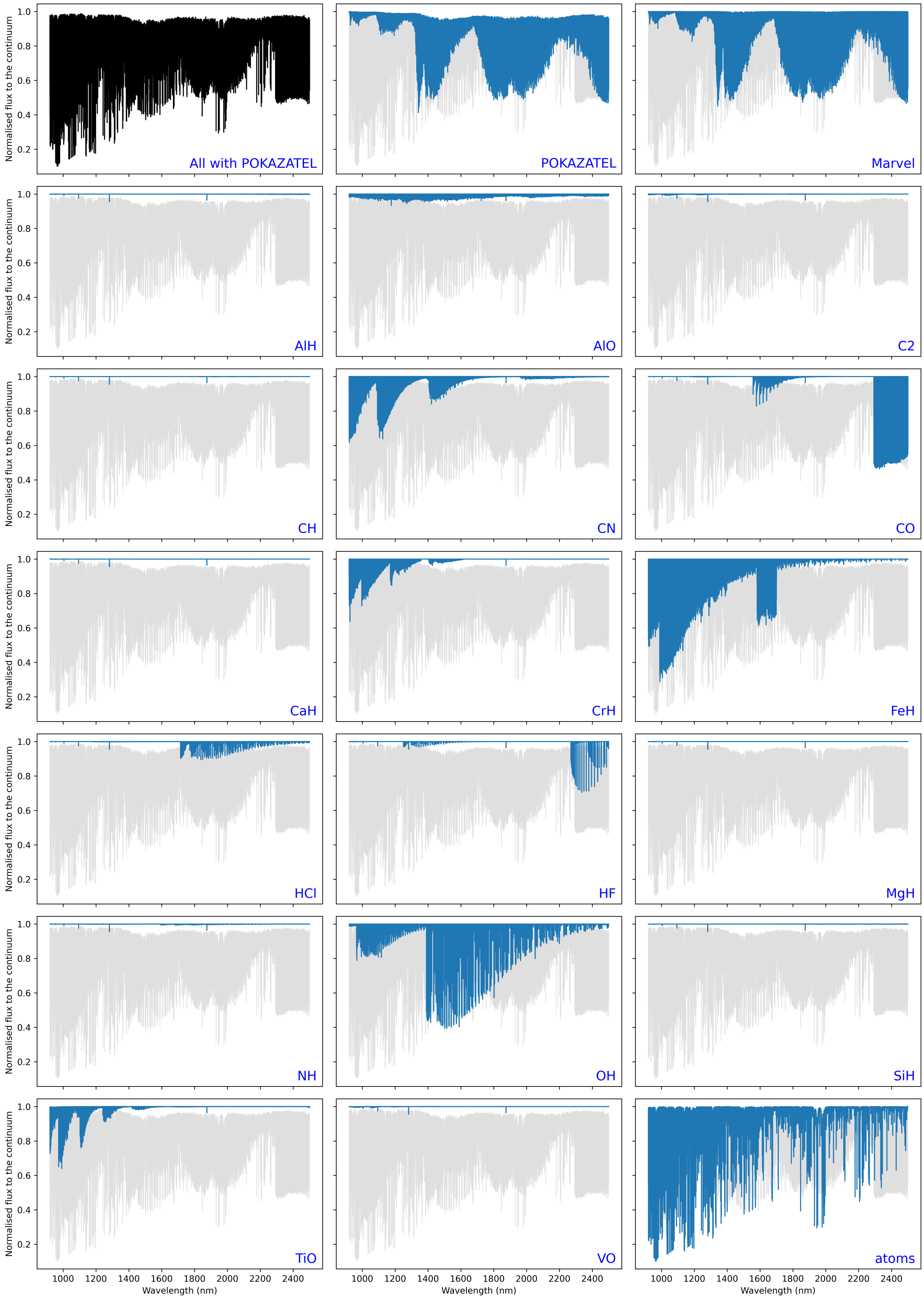}
      \caption{Continuum normalised synthetic spectra of M-dwarf star computed for the 3D RHD model of Table~\ref{simus} with a resolving power of $R=200,000$. Top left panel shows the spectrum including all the molecular and atomic transitions of Table~\ref{molatoms} with POKAZATEL linelist for the H$_2$O. The other panels show the contribution of each element (blue) overplotted on the complete spectrum (gray). Four different H$_2$O linelists are also shown.}
         \label{figsspectraHband}
   \end{figure*}

The spectra of M-dwarf stars are dominated by a dense forest of molecular absorption features, both in the optical and in the infrared, where their cool atmospheres allow a wide range of molecules to form and persist. Figure~\ref{figsspectraHband} displays the continuum normalised M~dwarf spectra computed at a constant resolving power of $R=200,000$. Top left panel shows the contribution of all molecular and atomic lines in the Y, J, H and K bands from Table~\ref{molatoms}, showcasing the numerous molecular transitions that blend and produce highly structured continua which is different from 1.0. Moreover, the spectra exhibit substantial variation in line depth, driven largely by the high molecular opacity and the steep temperature gradients within their cool atmospheres. Figure~\ref{figsspectraHband} displays also that molecules like OH, H$_2$O, FeH, and CO are largely dominating the general shape at different wavelengths. 

H$_2$O line blends (first row) are also the cause of the expected continuum, while OH has a visibility spectral window close to 1500 nm and CO at wavelengths larger than 2300 nm. Eventually, atomic lines are also visible but somewhat merged with molecular transitions for wavelengths larger than about 1200 nm. This molecular complexity makes the infrared spectra both rich in diagnostic potential and challenging to model, requiring high spectral resolution observations and sophisticated atmospheric opacities to accurately interpret stellar parameters, chemical abundances, and the environments of orbiting exoplanets.

\subsection{From one- to three-dimensional models: determination of micro- and macroturbulence}\label{micromacrosection}

Radiation hydrodynamics simulations provide a self-consistent ab initio description of the non-thermal velocity field generated by convection and overshoot that are only approximated in 1D hydrostatic 
models by empirical calibrations. Thus, the comparison between 1D and 3D spectra requires the determination of ad-hoc parameters like microturbulence ($\xi_{\rm{mic}}$) and macroturbulence ($\xi_{\rm{mac}}$) velocities that must be determined for the 1D spectra. To determine these parameters, we used classical 1D, LTE, plane-parallel, hydrostatic MARCS model atmosphere \citep{2008A&A...486..951G} with identical stellar parameters, chemical compositions, and linelists as the 3D simulation (Table~\ref{simus}). It should be noted that the MARCS models do not have identical input data and numerical scheme as the 3D simulation but they are a good benchmark for a quantitative comparison. 

\begin{figure}
   \centering
   \includegraphics[width=\hsize]{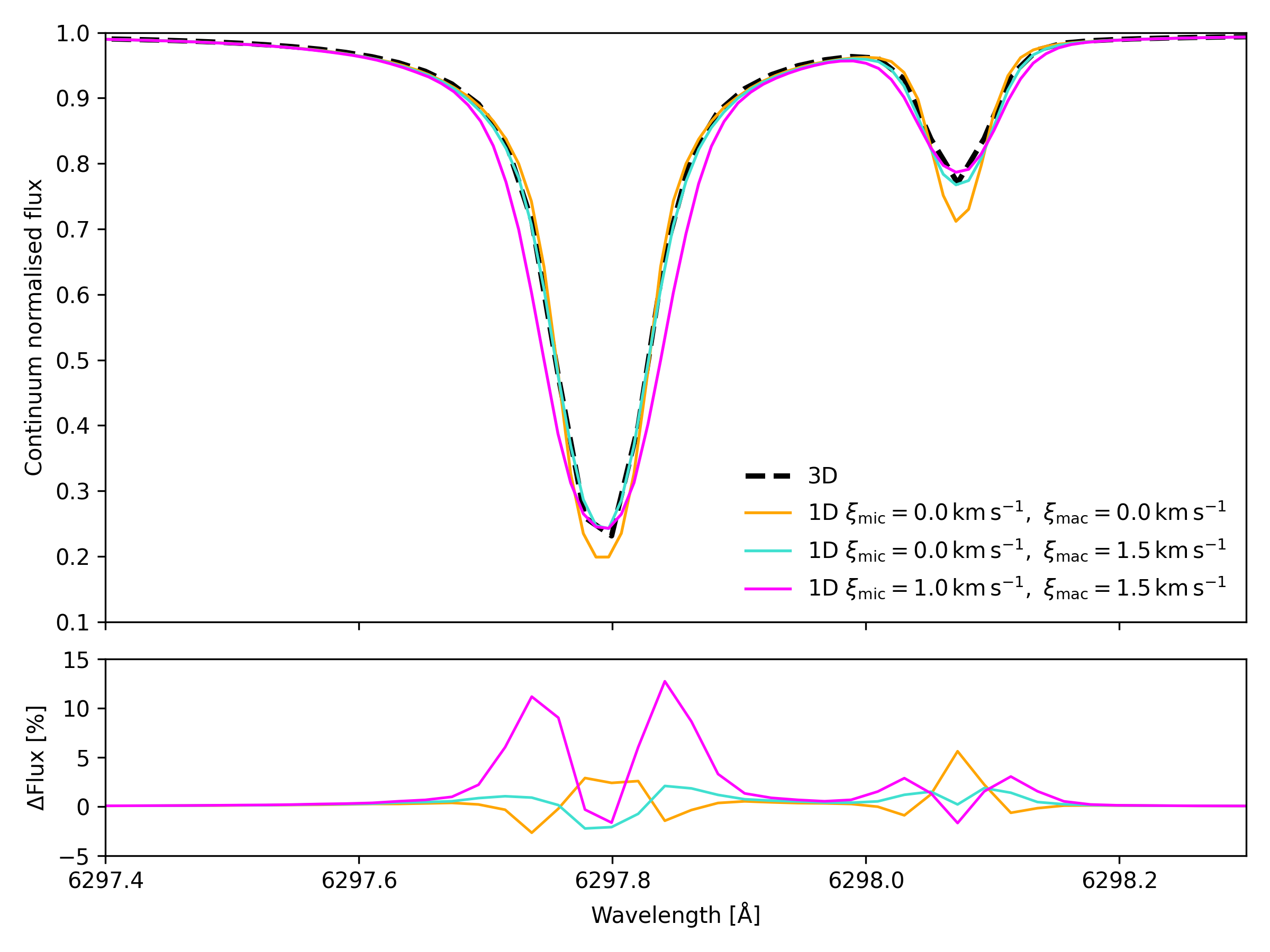}
      \caption{An example of micro- ($\xi_{\rm{mic}}$) and macroturbulence ($\xi_{\rm{mac}}$) fitting of 3D spectrum with 1D model of Table~\ref{simus} with different turbulence parameter realisations. The spectral range displays two representative iron lines among the 172 optical Fe~{\sc i} lines considered (see text).}
         \label{figmicromacro}
   \end{figure}

Standard 1D spectra usually include fixed $\xi_{\rm{mic}}$ and $\xi_{\rm{mac}}$ broadening that isotropically act on spectral lines, independently of depth, and identically for both atomic and molecular lines. Following a similar approach done in \cite{2009MmSAI..80..731S} and \cite{2011A&A...535A..22C}, we aim at achieving a good representation of the 3D spectrum using 1D spectra with different values of  micro-, and macro-turbulent velocity parameters. To do that, we used the 1D, pure LTE, \texttt{BACCHUS} code \citep{2016ascl.soft05004M} to compute 1D spectra for a range of 172 optical Fe~{\sc i} lines (Table~\ref{molatoms}) using the radiative transfer code Turbospectrum \citep{2012ascl.soft05004P}, and calculate the relative abundances with different $\xi_{\rm{mic}}$ and $\xi_{\rm{mac}}$ parameters. Figure~\ref{figmicromacro} displays an example comparison between 1D and 3D spectra. The microturbulence velocity is derived by requiring no trend between the abundances of the set of Fe~{\sc i} lines and their equivalent widths. \texttt{BACCHUS} also automatically derives the macroturbulence velocity by convolving the 1D synthetic spectra with a mean radial, tangential kernel to fit the Fe line profiles. Figure~\ref{figmicromacro} shows that the smallest difference between the 3D and 1D spectra is obtained for the 1D model computed with $\xi_{\rm mic} = 0.0$ km\,s$^{-1}$ and $\xi_{\rm mac} = 1.5$ km\,s$^{-1}$.
Using the entire set of Fe lines, we find that the microturbulence velocity required to reproduce the 3D spectra is lower than 0.1 km\,s$^{-1}$, while the macroturbulence velocity is $1.50 \pm 0.15$ km\,s$^{-1}$. The quoted uncertainties are derived from the line-to-line rms dispersion.

\subsection{Comparison with observations}\label{sec:model-data_comparison}

For this section, we used the 3D synthetic spectra computed for the M~dwarf simulation of Table~\ref{simus}. We used observations of GJ\,3470 (Table \ref{tab:gj3470}) obtained with the GIANO-B spectrograph \citep{Origlia2014} mounted at the 3.6-m Telescopio Nazionale Galileo (TNG). GIANO-B can observe simultaneously the whole near-infrared (0.95--2.45 $\mu$m) at a resolving power of $R=50,000$ and with sampling of $\sim2.7$\,km\,s$^{-1}$ per detector pixel. Thus, albeit at lower $R$, its spectral coverage is similar to the expected range of the ANDES instrument at the ELT, allowing us to test the model-data agreement in several near-infrared bands.

GJ 3470 is an excellent test case for these comparisons. It is a relatively bright ($K_s$ = 7.99 mag), early M1.5 dwarf star with a nearby point in the  \texttt{ Stagger}-grid, and is a slow rotator \citep[$v\sin i = 1.1\pm0.1$ km s$^{-1}$, derived from][]{Kosiarek2019}.
The presence of a Neptune-size companion \citep{Bonfils2012} makes the system even more interesting, and indeed there is a long history of studies dedicated to determine the system architecture, and the atmospheric composition and structure of the planet. GJ\,3470\,b belongs to a growing family of moderately eccentric warm Neptunes on a polar orbit \citep{Kosiarek2019, Stefansson2022} and is known to have an escaping atmosphere \citep{Bourrier2018, Palle2020}. Its primary atmosphere was also detected from space, first with HST \citep{Benneke2019} and then with JWST \citep{Beatty2024}, although the physical interpretation from the two datasets remains incompatible, especially regarding the role and scattering properties of a cloud layer. From the ground, the object does not have a detection yet \citep{Dash2024, Masson2026}, but it is considered a benchmark for the ability of high resolution spectroscopy to measure the spectrum above cloud layers \citep{Gandhi2020}. All in all, the interest of the community for this object makes it an ideal test case for our stellar modelling.

The spectra used here were obtained on January 27, 2020, in the context of the Global Architecture of Planetary Systems programme \citep[GAPS,][]{2017EPJP..132..364C}. They were subsequently extracted using the \texttt{GOFIO} v3.0 pipeline \citep{Rainer2018}, resulting in a time-series spanning from pre- to post-transit of the exoplanet GJ\,3470 b (from orbital phase $-0.017$ to 0.027, with phase zero corresponding to mid-transit). For the comparisons shown in this work, all the spectra were normalized using their median value across each spectral order, and then a subsequent median average across time was calculated, resulting in one single time-averaged spectrum per order. While the time average does contain the smeared exoplanet signature as well, its wavelength dependent transmission spectrum is expected to be $10^3$--$10^4$ times smaller than the stellar signal, hence it does not significantly alter the stellar spectrum.

\begin{figure*}
   \centering
    \begin{tabular}{c}
   \includegraphics[width=1.\hsize]{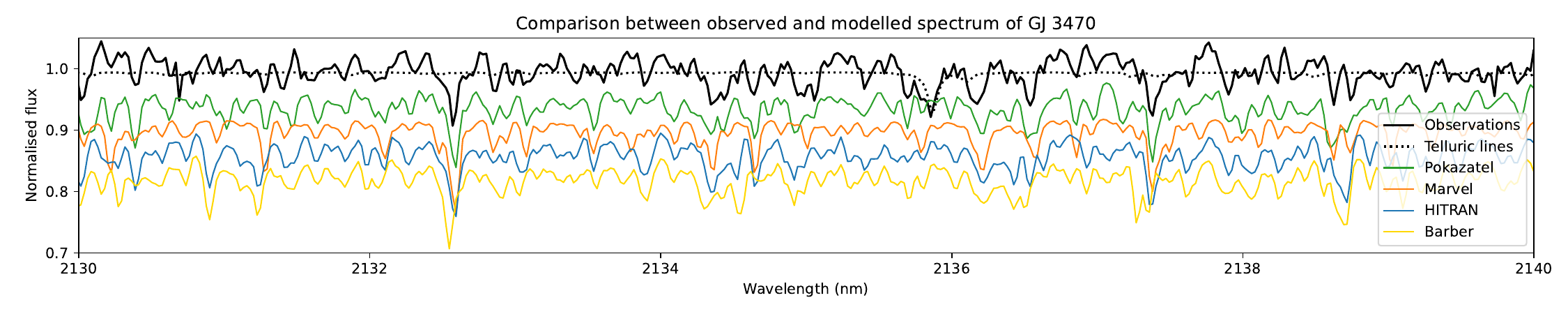}\\
         \end{tabular}
      \caption{Observed stellar spectrum of the M dwarf star GJ 3470 (black) in a spectral region dominated by H$_2$O lines. Telluric lines have been highlighted in dotted line. The different colored spectra correspond to calculations done with different linelists for H$_2$O (Table~\ref{molatoms}). Their pseudo-continuum has been artificially shifted to enable an easier comparison.}
         \label{figcomparisonobs1}
   \end{figure*}

We used the GIANO-B observations as a benchmark for the 3D spectra in the infrared range (Fig.~\ref{figsspectraHband}) to test the validity of the different linelists used (Table~\ref{molatoms}).
Figure~\ref{figcomparisonobs1} shows a comparison in a water dominated region between the observed data (black) and synthetic spectra (colors) computed with different H$_2$O lists. A precise comparison between model and observations is difficult to perform because several observed lines are shifted or even missing in the synthetic comparison.

\begin{figure}[h!]
   \centering
    \begin{tabular}{c}
      \includegraphics[width=0.9\hsize]{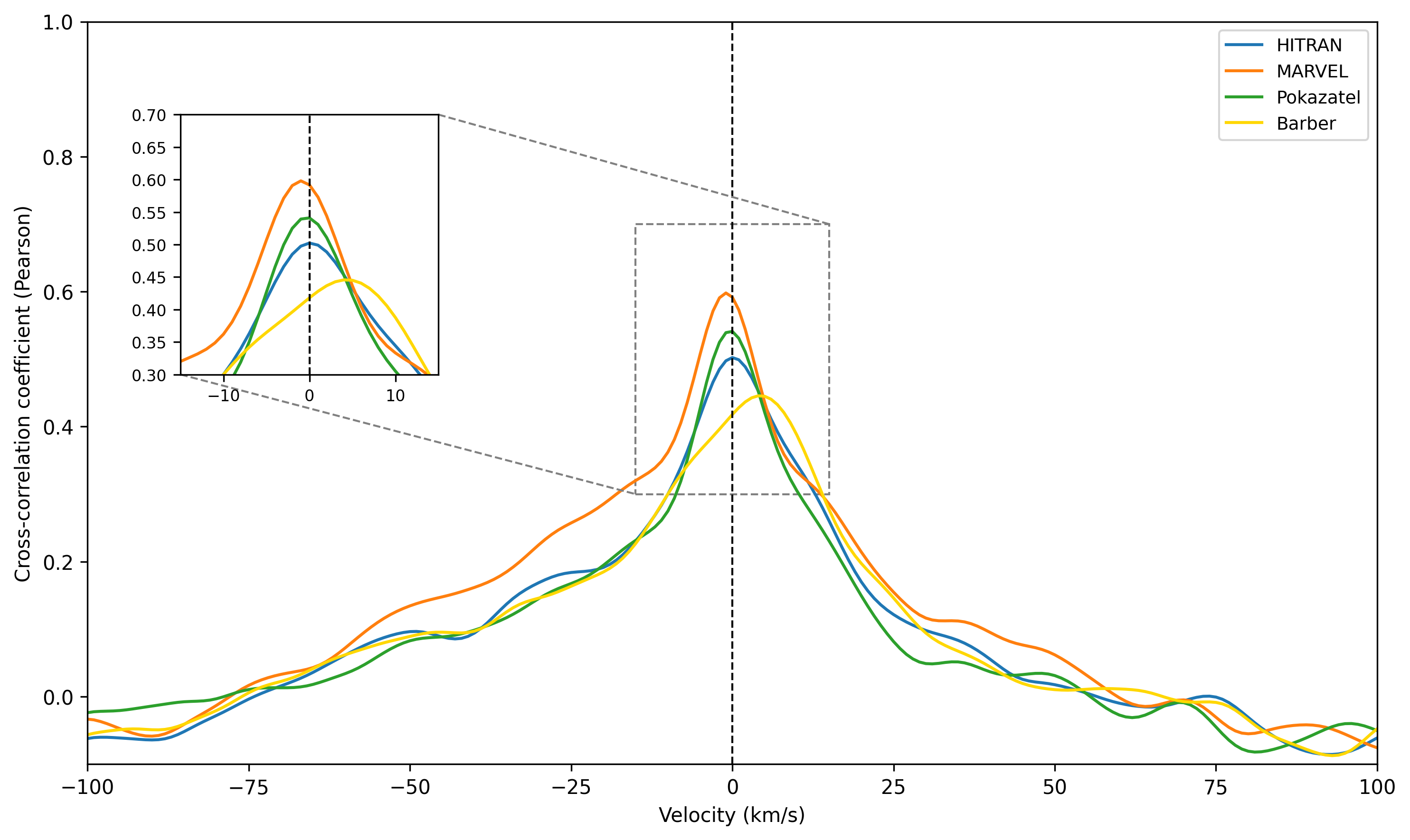} \\
       \includegraphics[width=0.9\hsize]{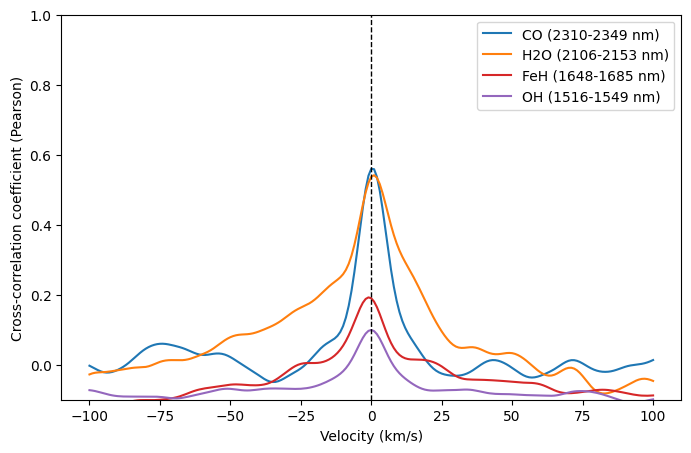}
       \end{tabular}
      \caption{\emph{Top panel:} Cross-correlation functions (CCFs) between different 3D synthetic spectra computed with H$_2$O linelists outlined in Table~\ref{molatoms} and GJ 3470's observed spectrum (as shown in Fig.~\ref{figcomparisonobs1}). The peak position of the MARVEL CCF is at -1.1 km\,s$^{-1}$, POKAZATEL at +0.1 km\,s$^{-1}$, HITRAN at +0.78 km\,s$^{-1}$, and Barber at > +5.0 km\,s$^{-1}$. \emph{Bottom panel:} Cross-correlation functions computed in different spectral regions where CO, FeH, and OH are the principal contributors to the absorption spectrum. For calculation with H$_2$O, we use the spectrum with the POKAZATEL line list.}
         \label{figcomparisonobs2}
   \end{figure}

To quantify the goodness of the water linelists, we cross-correlated the different spectra with the GJ 3470 data in the H$_2$O-related spectral range. Prior to cross correlation, the model spectra were broadened with a  Gaussian kernel of FWHM = 6 km s$^{-1}$ to match the $R=50,000$ of GIANO-B. Figure~\ref{figcomparisonobs2} (top panel) shows that the cross-correlation coefficient is lower than 0.6 for all four lists. This means that they are missing molecular/atomic species necessary to recover all the observed depth and shape of the spectral lines. It should be noted that we selected this region because it is (almost) telluric-free (dotted line in Fig.~\ref{figcomparisonobs1}). While the MARVEL list returns a higher coefficient ($\approx$0.6), the cross-correlation function shape is not centered at 0 km\,s$^{-1}$. HITRAN and POKAZATEL have a lower coefficient ($\approx$0.5--0.55) but their peak is centered at 0 km\,s$^{-1}$ (POKAZATEL being the closest). Finally, the Barber list appears highly asymmetric, with a lower coefficient ($\approx$0.45), and largely diverging from 0 km\,s$^{-1}$. We note that absolute RV shift should be taken with a grain of salt, because GIANO's wavelength solution is typically only accurate at the 0.2--0.5 km\,s$^{-1}$ level.
Therefore, relative RV shifts between datasets are more informative than absolute RV shifts from zero at this stage. \\

The peak position of the MARVEL CCF is at -1.1 km\,s$^{-1}$, for the POKAZATEL +0.1 km\,s$^{-1}$, and for HITRAN +0.78 km\,s$^{-1}$. However, the differences between the POKAZATEL, MARVEL, and HITRAN linelists remain relatively small, making it difficult to identify a clearly superior dataset. In the case of the Barber list, the shift is much larger, peaking at values larger than +5.0 km\,s$^{-1}$. The MARVEL list is effectively a subset of POKAZATEL, in which the energy levels have been empirically adjusted and, in principle, refined. The two lists differ both in the total number of transitions and in the resulting continuum placement.
Nevertheless, none of the adopted line lists fully reproduces the observed spectrum. The incomplete coverage of weak and blended features constitutes a limiting factor in the final stellar correction and, consequently, it may impact the robust identification of water signatures in exoplanet atmospheres. Refining the stellar metallicity in the 3D simulations and extending the comparison to a broader sample of stars spanning different stellar parameters would significantly strengthen and improve the analysis.

Figure~\ref{figcomparisonobs2} (bottom panel) shows the cross-correlation functions computed in additional spectral regions, compared to the one obtained for water. Based on Fig.~\ref{figsspectraHband}, we selected a CO-dominated region (2310--2349 nm), an FeH-dominated region (1648--1685 nm), and an OH-dominated region (1516--1549 nm).
The CO cross-correlation function is the narrowest and peaks at 0 km\,s$^{-1}$, while the FeH and OH regions yield lower correlation coefficients. Telluric contamination and/or instrumental systematics may also affect this comparison and could, in particular, contribute to reducing the amplitude of the CO cross-correlation signal.

\section{Comparing stellar time-variable effects to an exoplanet transmission spectrum}\label{sec:clv+rme_model}

Besides allowing a direct validation of the stellar modelling as described in previous section, 
it is informative to simulate the GJ\,3470 system to understand whether the lack of correction of stellar effects in transmission in published literature could be the cause of the non-detection at high resolution. In general, warm Neptunes or sub-Neptunes around M dwarfs are an interesting case study from the point of view of planetary architectures, in that many of them tend to be on misaligned orbits and with small but non-negligible eccentricity, which might point to early dynamical scattering of the planet from an outer orbit to their final, current close-in configuration \citep{2018Natur.553..477B}.

\begin{table}
\tiny
\centering
\begin{minipage}[t]{\columnwidth}
\caption{Parameters of the system GJ 3470}             
\label{tab:gj3470}      
\centering                          
\renewcommand{\footnoterule}{} 
\begin{tabular}{lll}        
\hline\hline                 
Parameter & Value & Reference \\
\hline
Stellar radius ($R_\star$) & $0.48\pm 0.04\ R_\odot$ & \citet{Kosiarek2019} \\
Effective temperature ($T_\text{eff}$) & $3650 \pm 50$ K  & \citet{Kosiarek2019} \\
Rotational period ($P_\text{rot}$) & $21.5\pm0.5$ d & \citet{Kosiarek2019} \\
Rotational velocity ($v\sin i$) & $1.1\pm0.1$ km\,s$^{-1}$ & this work \\
Metallicity ([Fe/H]) & $0.2\pm0.1$ dex & \citet{Kosiarek2019} \\
$\log{g}$ & $4.66\pm0.04$ cm\,s$^{-2}$ & \citet{Kosiarek2019} \\
\hline
Semimajor axis ($a/R_\star)$ & $13.9^{+0.4}_{-0.5}$ & \citet{Kokori23} \\
Inclination ($i_p$) & $88.9^{+0.6}_{-0.5}$ deg & \citet{Kokori23} \\
Impact parameter ($b$) & $0.267\pm0.034$ & this work \\
Planet-to-star radius $(R_\mathrm{P}/R_\star)$ & $0.0764\pm0.0004$ & \citet{Kokori23} \\
\hline                          
\end{tabular}
\end{minipage}
\end{table}

\subsection{Planet transit model}\label{sec:clv+rme_model}


The presence of an exoplanet in transit produces two types of distortions on the stellar lines, namely centre-to-limb variations (CLVs) and Rossiter-McLaughlin effect (RME). To correctly model both effects, for highly inclined orbits, we upgraded the code of \citet{2019A&A...631A.100C} to reproduce a tilted geometry for both the planet orbit and the stellar rotation axis. A list of system parameters used in this work is provided in Table \ref{tab:gj3470}, along with the literature sources. 
Details of the code upgrades are provided in Appendix \ref{sec:appendix_clv-rme_model}, while here we recap the fundamental working principles.

The stellar surface is represented with a Cartesian grid with a fixed number of pixels (200 per axis in these simulations). Each pixel has a velocity field and a $\mu$ value, where the latter is computed as in \citet{2019A&A...631A.100C} while the former is given by Eq. \ref{eq:v_LOS}. Each pixel also has a weight attached, which is 1 when the stellar disk entirely fills the pixel, and between 0 and 1 for pixels only fractionally occupied by the stellar disk. Partial occupancy is computed by oversampling the stellar disk by a factor of 25.

As a first step, the normalized, disk integrated stellar spectrum is computed by summing up the contribution of each pixel, i.e. by linearly interpolating the 3D model grid presented in Sect.~\ref{sec:spectra} to match the correct tilt $\mu$, and then by spline interpolating in wavelength to apply the correct Doppler shift. Each pixel is also weighted via its fractional occupancy computed as explained in the previous paragraph. In the second step, the occlusion of the planet disk is calculated from the orbital phase, semimajor axis, and impact parameter of the system. As for the stellar disk, pixels fractionally occluded are also assigned a weight to reduce errors coming from the relatively low number of pixels describing the planet disk. In this case, the oversampling factor is 100 on each axis, obtained for computational reasons by extracting 10,000 combinations of ($x,y$) subpixel coordinates from a uniform distribution rather than from a 100$\times$100 regular grid. In the last step, rather than recomputing the full disk stellar spectrum with the occluded pixel excluded, we subtract from the disk integrated spectrum at each orbital phase the ``spectrum behind the planet'', i.e. the contribution of each occluded pixel, again accounting for its line of sight velocity and tilt angle with the double interpolation explained above.

We estimated the impact of the grid size on the precision of the simulations. We set 500 pixels per stellar radius as our benchmark, and then tested values of 10, 20, 50, 100, 250 pixels per stellar radius. We found that 100 pixels per stellar radius (i.e. a $200\times200$ pixel grid for the whole star) is the best compromise between speed and precision. It corresponds to a relative precision of 5$\times10^{-6}$ in the computation of the disk-integrated stellar flux, and $4\times10^{-8}$ in the phase resolved CLV + RME distortions of the spectrum. Interpolation inaccuracy from choosing spline vs linear interpolation is below the above level of precision in our tests, but we nevertheless opt for spline interpolation to better capture the smooth, noiseless nature of the spectral models.

The output of our code is a stellar model matrix which has wavelength on the horizontal axis and time on the vertical axis, each of the latter corresponding to a planetary phase. While these models by construction include CLVs and RME, they do not include by default systemic velocity, barycentric velocity correction, and any stellar reflex motion due to the presence of the planet, the reason being that these quantities are often unknown at the level of precision required by the subsequent correction (due to actual measurement uncertainty, but also wavelength calibration errors in many of the slit-fed infrared spectrographs), requiring subsequent fine tuning via the application of per-spectrum Doppler shift via spline interpolation, as described e.g. for HD\,189733\,b in \citet{Flowers2019} and \citet{2019A&A...631A.100C}. We therefore leave to the user to optimize / include these further parameters via subsequent interpolation of our output model.

\begin{figure*}
   \centering
   \begin{tabular}{cc}  
        \includegraphics[trim={0 0 3.2cm 0},clip,width=0.4\hsize]{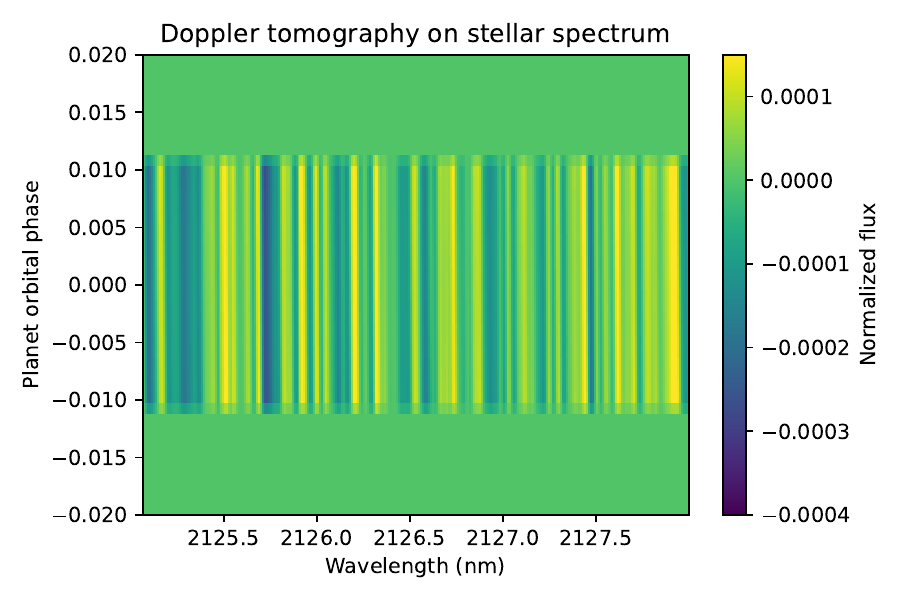} 
        \includegraphics[trim={2.3cm 0 0 0},clip,width=0.43\hsize]{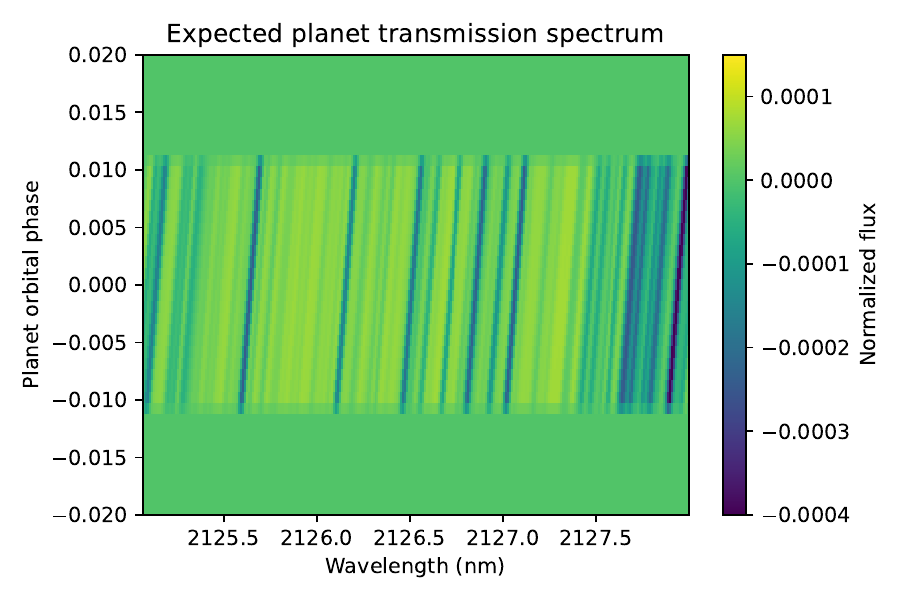} 
    \end{tabular}
    \caption{\textit{(Left panel):} Doppler tomography of the stellar model of GJ 3470 distorted by the transit of planet 'b', as explained in Sect.~\ref{sec:clv+rme_model}. \textit{(Right panel):} time resolved planetary transmission spectrum of GJ 3470 b with the calculation as explained in Sect.~\ref{sec:svd_performances}. Both spectral sequences are normalized in units of stellar flux.}
    \label{dopplerstar_expect_planet}
\end{figure*}

As an additional note, our stellar models also output the continuum for each of the wavelength channels. Here we use such continuum to normalize the stellar spectrum in output of the CLV+RME model. The main distortions of the stellar lines can be effectively visualized by applying Doppler tomography on the output model matrix via the following procedure. Firstly, a master out-of-transit spectrum is computed by averaging all the out-of-transit spectra in time. Each spectrum in the sequence is then divided by the master, removing the mean undistorted stellar profile. As a second step, a white light curve is constructed by averaging in wavelength all the spectra normalized as above. Each spectral channel (i.e. each wavelength) is then divided by the light curve to remove the effect of the planetary transit. The final product is shown in Fig. \ref{dopplerstar_expect_planet} (left panel), compared to the best-fit JWST transmission spectrum of \cite{Welbanks2024} (right panel), Doppler shifted and scaled as explained in Sect.~\ref{sec:svd_performances}. We note that the simulations are computed across the spectral range 2125--2145 nm, but in Fig. \ref{dopplerstar_expect_planet} and subsequent figures we only visualize selected sub-ranges to properly highlight the spectral components or their time variability.

\subsection{Using principal component analysis to de-trend stellar effects}\label{sec:svd_performances}

Figure \ref{dopplerstar_expect_planet} shows that for a misaligned system like GJ 3470 and due to the slow stellar rotation the combined CLVs+RME is significantly more stationary than the exoplanet spectrum. Moreover, as expected, 
the sign of the spectral features is ``flipped'' (i.e. excess residual in the stellar case, absorption in the planetary case). Nevertheless, the amplitude of the stellar effect is of the same order of magnitude as the exoplanet signal, and thus the effect might not be negligible.
In common analyses of near-infrared spectra, principal component analysis \citep[PCA,][]{deKok2013} or \texttt{Sysrem} are used to remove stationary or quasi-stationary spectral features (usually the Earth's atmospheric absorption and the out-of-transit stellar spectrum) while preserving a large fraction of the Doppler-shifting exoplanet spectrum.
Therefore, it is interesting to investigate whether PCA could effectively de-trend the stellar CLVs+RME, and we provide here a first quantitative estimate.

\subsubsection{Modeling a realistic observing sequence}\label{sec:tell_sim}

In this exploratory analysis, the exoplanet transmission spectrum is merged with the CLV+RME model in a straightforward way. We use the 1D radiative transfer model shown in Fig. \ref{dopplerstar_expect_planet} matching the JWST detection of GJ 3470 b's atmosphere from \citet{Welbanks2024} and computed with \texttt{GENESIS} \citep{Gandhi2019} as the planet's transmission coefficient $T_p$:
$$T_p(\lambda) = 1 - s\left(\frac{R_p(\lambda)}{R_\star}\right)^2,$$
with $s$ a scaling factor defined further below. We add the temporal dimension by Doppler shifting the planet by its observed radial velocity $v_P$ as a function of time $t$:
$$ v_P(t) = V_\mathrm{sys} + V_\mathrm{bary}(t) + K_P\sin[2\pi\varphi(t)],$$
where $\varphi$ is the planet's orbital phase, $V_\mathrm{sys}$ and $V_\mathrm{bary}$ the systemic and barycentric velocities, and $K_P$ is the planet's radial velocity semi-amplitude, that is:
$$ K_P = v_\mathrm{orb} \sin i_P = \frac{2\pi a}{P}\sin i_P,$$
with $a$ the semi-major axis, $P$ the orbital period, $i_P$ the orbital inclination. At each orbital phase, the planet is multiplied by a scaling factor $s$ obtained by averaging in wavelength the CLV+RME simulations, multiplying by $-1$ the resulting light curve and normalizing it so that the out-of-transit portion weights zero and the centre of transit weights 1. 

\begin{figure}
   \centering
   \includegraphics[width=0.85\hsize]{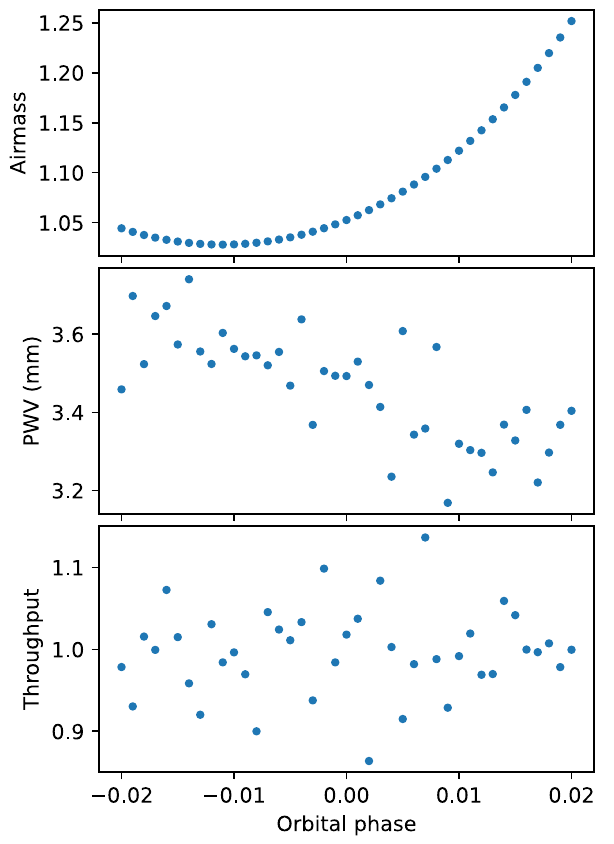}
   \caption{From top to bottom: airmass curve, precipitable water vapour and throughput variations adopted in the simulations described in Sect.~\ref{sec:tell_sim}.}
   \label{fig:sim_pars}
\end{figure}

For a realistic simulation of the PCA behaviour, the Earth's transmission spectrum needs to be input realistically, at least by including airmass and PWV variations. We do so via bi-linear interpolation in log-transmission over a precomputed grid of telluric spectra via the ESO Sky Model Calculator, and then via airmass scaling with a simple Beer-Lambert law. We also include throughput variation due e.g. to pointing inaccuracy and seeing variation by modulating the overall flux with random scaling factors drawn from a Gaussian distribution centered at 1.0 and with a standard deviation of 0.05 ($\pm5\%$ variability). Airmass and PWV are adapted from the actual observing night with GIANO described in Sect.~\ref{sec:model-data_comparison}, and the simulation parameters are visualized in Fig.~\ref{fig:sim_pars}.
To obtain the noiseless combined spectrum as a matrix with dimensions $(\lambda,t)$ we multiply the three matrices of the CLV+RME effect, variable telluric transmission, and variable exoplanet transmission.

White noise is added by assuming a bright-source limit where the photon counting from the host star is the dominant source of (photon) noise. The spectra are normalized, and a given signal-to-noise ratio (S/N) corresponds to a normalized flux of 1.0. The S/N per pixel again in a $(\lambda,t)$ two-dimensional representation is obtained via S/N times the square root of the normalized flux.
Random Gaussian noise centred at zero and with standard deviation 1/[S/N$(\lambda,t)$] is then 
added to the normalized flux. 



\subsubsection{PCA detrending}\label{sec:PCA}

Our aim here is to show whether PCA can de-trend stellar effects, and whether the behaviour of PCA is influenced by the presence of multiple time-variable effects and noise. To do so, we process the spectral sequences (i.e. the simulated observations) described above in the time domain, i.e. considering time as the independent variable, and exclusively via Singular Value Decomposition \citep[SVD, implemented via \texttt{numpy.linalg.svd()} as in][]{deKok2013, Line2021} rather than with the additional standardization of \citet{Giacobbe2021}. We only show and comment on the first four SVD components, but we have verified that in this specific case further components do not change the overall qualitative behaviour of the analysis.





\begin{figure*}
    \centering
    \includegraphics[width=0.95\hsize]{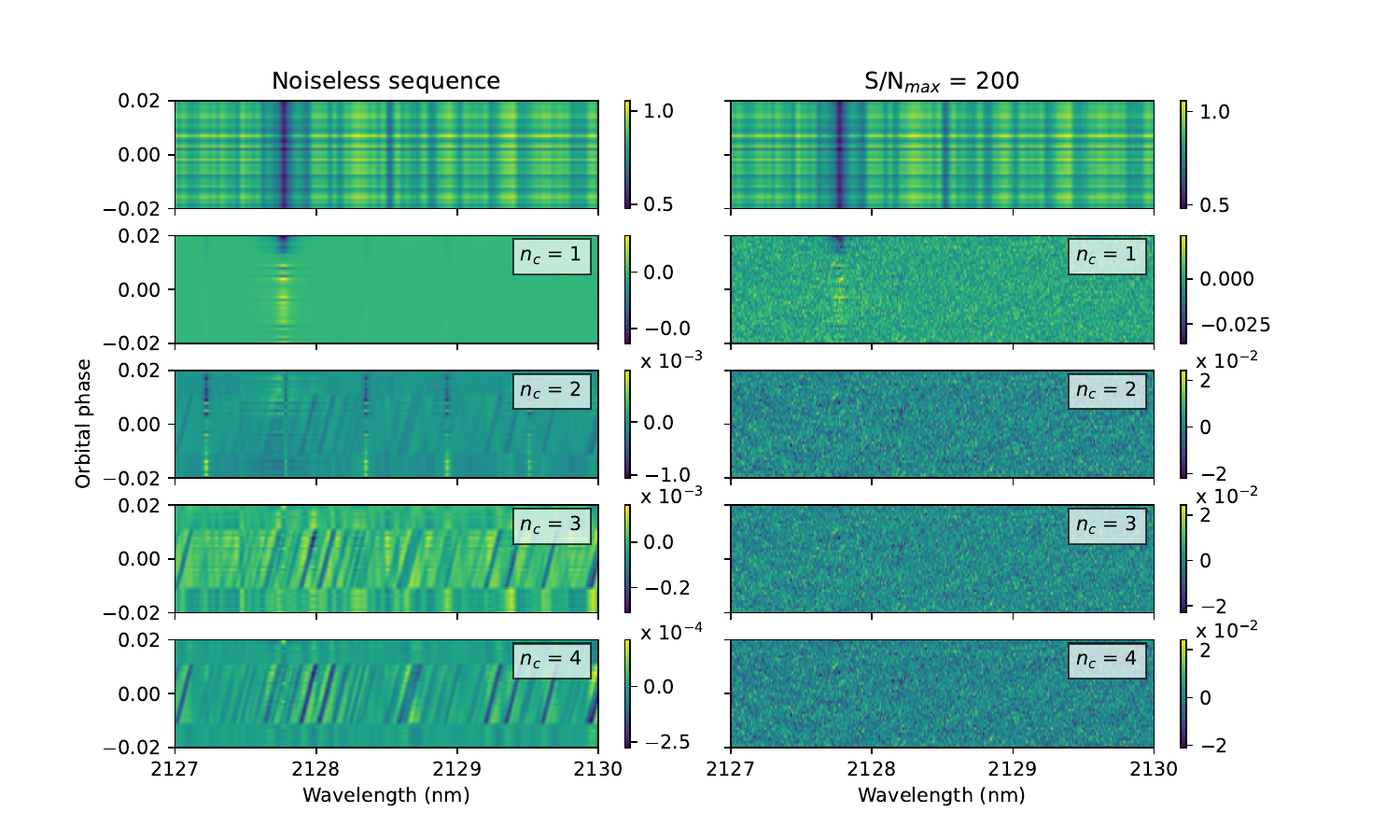}
    \caption{Performances of Singular Value Decomposition (SVD), as described in Sect.~\ref{sec:PCA}, in removing astrophysical time-variable components. The noiseless (left panel) and noisy (right panel) datasets are shown, both at the initial stage (top row) and for progressively higher numbers of SVD components ($n_c$) removed (2$^{nd}$ to 5$^{th}$ row).}
    \label{fig:pca_residuals}
\end{figure*}

In Fig.~\ref{fig:pca_residuals} we show the simulated observing sequence (top row) in the noiseless and S/N = 200 cases (left and right column, respectively), as well as the residual cleaned matrices removing a progressively higher number of components (subsequent rows). These can be obtained in the SVD decomposition by setting to zero the first $n_c$ singular values, where $n_c$ is the number of components one wishes to remove.
In the noiseless case (left panels) it is possible to appreciate that SVD first tackles the main airmass + throughput variations ($n_c=1$), then focuses on residual water telluric variations ($n_c=2,3$) and finally de-trends the CLV+RME effects ($n_c\ge4$), with a level of success that is not easy to evaluate visually. We note that the bottom-left panel shows the typical alterations that PCA/SVD imprints on a real exoplanet signal, with distorted lines and altered amplitudes (compare, e.g., with the original exoplanet spectrum shown in Fig. \ref{dopplerstar_expect_planet}, right panel). 
In the presence of noise (right column), at least visually, the same SVD reconstruction flattens after only 2 components. We know from the left panel that there is underlying residual correlated noise, but such correlated noise is not visible at the level of the simulated data, because it is buried in the noise. This fact is quite concerning, because the two most common practices to stop SVD (i.e., to select the optimal $n_c$) are based either on visual inspection or on the rate of relative change in the standard deviation of the residuals. Both criteria would indicate $n_c$=2 as the optimal number of SVD components in this case, which would leave subtle telluric residuals (the regularly spaced vertical lines in the middle row -- noiseless).

\begin{figure*}
    \centering
    \includegraphics[width=0.95\hsize]{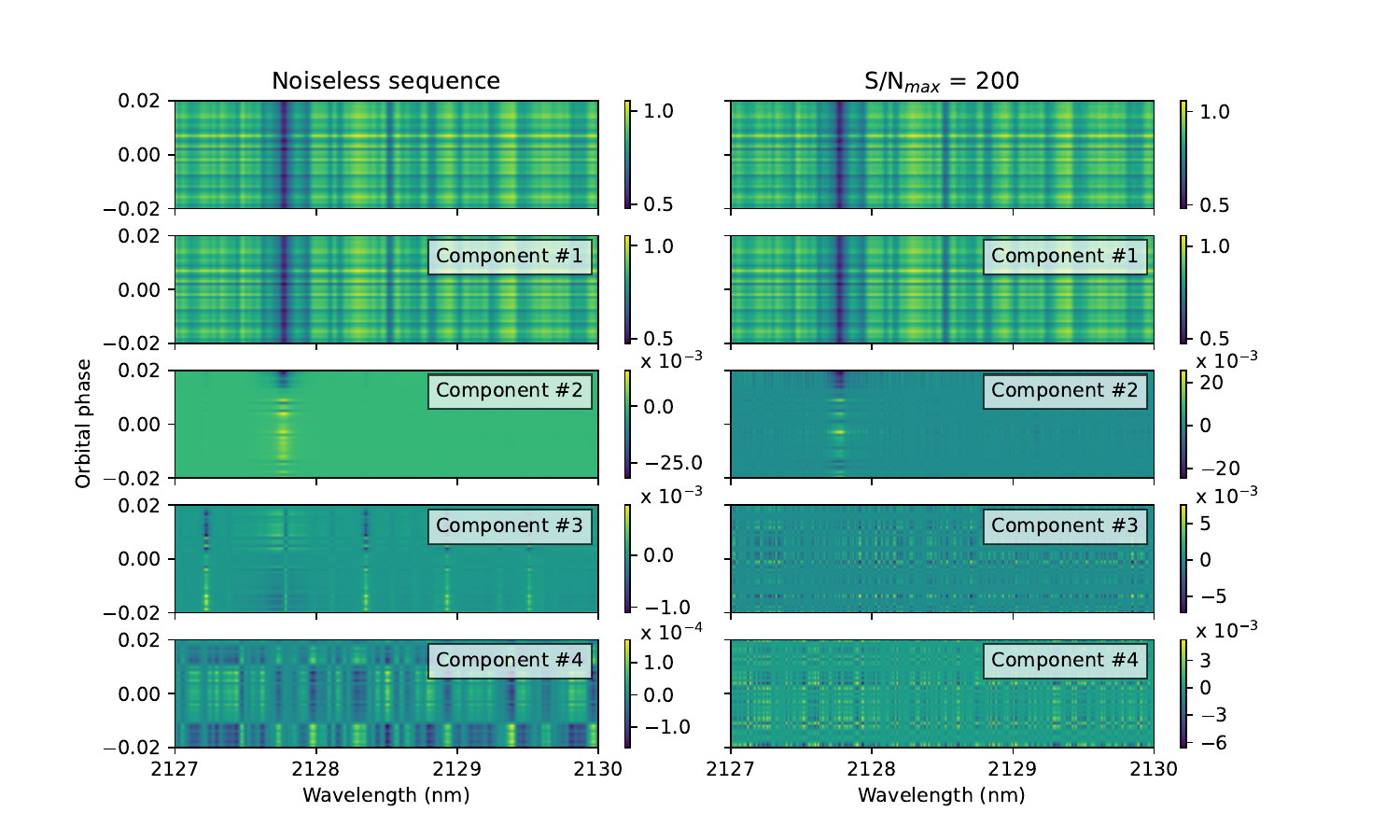}
    \caption{Similar to Fig. \ref{fig:pca_residuals}, except that we show each component reconstructed by the SVD algorithm sequentially in each row. See Sect.~\ref{sec:PCA} for details.}
    \label{fig:pca_components}
\end{figure*}

In light of the above, it is reasonable to ask ourselves whether SVD actually finds different components (different singular vectors) in the presence of noise. Figure~\ref{fig:pca_components} visualizes this point by showing the reconstructed components for $n_c$=1 to 4. These are obtained by zeroing all the singular values except for the $n_c$-th, and then by reconstructing the matrices as usually done in SVD.
In the noiseless case, our per-component visualization confirms that indeed components 1 to 3 model throughput and telluric variations, whereas component 4 starts modelling trends related to the transit of the planet and possibly the CLV effect. 

In contrast, in the S/N = 200 case (right column) the situation is remarkably different, with the noise floor starting to influence the components found via SVD starting from $n_c=3$. We can see that while subsequent components must be different by construction (SVD is a closed linear algebra operation with orthogonal singular vectors), their association with actual physical effects is impossible and the trends seem to follow large scale variations which might as well be associated with noise modulations. We also note that the amplitude of these reconstructed components is typically a few 10$^{-3}$, which is comparable to the noise per pixel ($5\times10^{-3}$ at S/N = 200). This shows that SVD uses the information from hundreds of spectral channels (i.e. hundreds of columns in our 2D representation) and therefore can reconstruct effects below the photon noise level, however the typical amplitude of astrophysical components (left panel, bottom rows) is of the order of 10$^{-4}$ i.e. approximately another order of magnitude below the deepest SVD reconstruction. Interestingly, OH is another species expected to be present in M-dwarf atmospheres which could interfere with similar searches in exoplanet atmospheres. Indeed, in the GJ\,3470 observations of \citet{Masson2026}, an anomalous, tentative signal from CH$_4$+OH is detected at low $K_P$. While lacking the typical V-shaped flaring reported in \citet{2019A&A...631A.100C}, it is possible that the application of \texttt{Sysrem} to clean the data, similarly to SVD in this study, cannot entirely remove correlated signals significantly below the noise level per-spectrum. This possibility further motivates follow up quantitative studies on the efficacy of data-driven de-trending in removing stellar contaminants.

As a further test, shown in Fig.~\ref{fig:svd_separate_effects}, we ran SVD in the noiseless case, but on the separate CLV+RME model and planet transit model (left and middle panels, respectively), and compare it with the noiseless decomposition already shown in Fig.~\ref{fig:pca_components} and merging all the effects together, including telluric and instrumental effects. It shows that the SVD components found are qualitatively very different, and thus we cannot assess the performances of SVD by modelling each component separately. On the contrary, only a full simulation including stellar effects, planet signal, instrumental effects, and noise modulation can predict how effective SVD actually is in detrending stellar effects. In the specific case of GJ 3470 b, even at a moderate S/N of 200 per wavelength bin (typical of 8-10\,m class telescopes), it is likely that SVD would fail at effectively removing the signature of CLV and RME, effectively requiring some prior modelling of the stellar effects.

Lastly, since Figs. \ref{dopplerstar_expect_planet} and \ref{fig:pca_components} show a similar amplitude between features in the exoplanet transmission spectrum and CLV+RME residuals, cross-correlating SVD residuals 
with exoplanet models
may still introduce spurious correlation from residual stellar effects. While a full Bayesian analysis to quantify potential biases in measuring abundances and temperatures is beyond the purpose of this study, we present in Appendix~\ref{sec:appendix_detection_tests} trail maps (CCF vs orbital phase and RV) and ($V_{rest}, K_P$) maps, commonly used to claim detections of exoplanet atmospheres. In this specific case, we do not measure any significant impact of stellar RME+CLV on the detection of the transmission spectrum of GJ 3470\,b at the correct velocity. We stress that whereas in this section we adopted S/N=200, typical for 8-10m class telescopes, in Appendix~\ref{sec:appendix_detection_tests} we need to boost the S/N to 500 to achieve a significant planet detection. We further comment in the Conclusions below about the implications of this choice. Furthermore, we also highlight that these preliminary simulations are limited by choice to a spectral range fairly devoid of telluric lines, and focussed on relatively weak stellar water lines, i.e. blue-ward of the potentially stronger signature stellar CO lines (the 2-0 ro-vibrational band onsets at around 2.29 $\mu$m).




\section{Conclusions}

In this paper, we have highlighted some of the challenges related to the 3D modelling of M-dwarf stars, with a particular focus on current inaccuracies in opacity line lists, and studied the nuisance effects.
We have presented the high-resolution ($\lambda/\Delta\lambda = 200,000$) synthetic spectra from a 3D RHD simulation covering both the visible and infrared spectral domains, including the $\mu$-dependent center-to-limb variation and a temporal sequence spanning approximately two hours of  convective surface evolution.
With these spectra we concluded: 

\begin{itemize}
\item The spectra of M-dwarf stars are shaped by an extremely dense forest of atomic and molecular lines, making them both highly informative and difficult to reproduce accurately. High-resolution synthetic spectra combined with detailed molecular opacities are therefore essential to reliably constrain stellar atmospheric properties and exoplanet environments.
\item 3D radiation-hydrodynamics simulations naturally reproduce the non-thermal velocity fields generated by convection and overshoot, avoiding the empirical approximations required in 1D hydrostatic models. This makes 3D stellar simulations also well suited for direct comparison with real spectroscopic observations.
The closest agreement between 1D and 3D spectra is obtained adopting $\xi_{\rm mic}$ = 0.0 km\,s$^{-1}$ and $\xi_{\rm mac}$ = 1.5 km\,s$^{-1}$ in the 1D calculations. 
\item The comparison with GIANO-B observations of GJ 3470 confirms that current H$_2$O molecular linelists still struggle to fully reproduce the observed infrared spectral features. Among the tested datasets, MARVEL provides the highest correlation coefficient, while HITRAN shows the best velocity alignment, highlighting the need for further improvements in molecular opacity databases for cool-star spectroscopy. In this context, M~dwarf stars are particularly well suited for being excellent laboratories for comparing synthetic spectra generated from different line lists with observed stellar spectra, because their cool atmospheres allow producing strong molecular absorption bands in both the optical and near-infrared regions \citep{2018A&A...620A.180R}.
\item Simulation of a full time-resolved exoplanet signal for the GJ 3470 system shows that typical application of PCA-based detrending methods cannot completely remove stellar time-correlated effects due to CLVs and RME. However, in this particular case, the detection of GJ 3470 b's atmosphere at the correct velocity does not seem affected. 
\end{itemize}

While the exoplanet transit simulations in Sect.~\ref{sec:clv+rme_model} are qualitatively solid, a more comprehensive quantitative assessment must address the following shortcomings:

\begin{itemize}
    \item The CLV+RME simulations need to be run for a variety of planet sizes, system obliquities, and stellar rotational rates, as the Doppler signature of these effects is strongly dependent on these parameters.
    \item The planet transmission spectrum can be inherently included in the stellar RME+CLV model as a chromatic effect by using a wavelength dependent planet radius $R_P(\lambda)$, although this currently produces a computational bottleneck as the simulation time would increase approximately by the number of spectral channels simulated, i.e. a factor of 10$^4$-10$^5$.
    \item SVD based detrending performed in the wavelength domain might perform differently compared to an application in the time domain, and since both approaches are present in the literature, a thorough comparison would be helpful.
    \item While the S/N of 500 assumed in Appendix \ref{sec:appendix_detection_tests} to recover GJ 3470\,b's signal may appear demanding, this is largely a consequence of the intentionally restricted spectral window (2125-2145 nm) used. Current high-resolution spectrographs cover much broader wavelength ranges, providing a substantially larger gain from cross correlation and thus reducing the S/N requirements. On the other hand, the broader spectral coverage introduces additional complexity, as the impact of telluric and stellar contamination will vary across different wavelength regions. Therefore, to
    completely estimate the actual impact of RME+CLVs on detection and characterization of exoplanet atmospheres, one would need to produce a full observation spanning the whole 1.0-2.4 $\mu$m range of current instrumentation at realistic S/Ns. Furthermore, the typical cross correlation detection matrices presented here should be paired with Bayesian retrievals to estimate biases in measured abundances and temperatures. 
\end{itemize}
While these caveats are arguably beyond the point of the current work, they will need to be addressed in future work to obtain a full quantitative indication of the best operative strategy to treat the copresence of stellar and planetary signals in high-resolution spectroscopy of planets around M-dwarf stars.

Our work highlights a fundamental limitation of SVD/PCA-based approaches: for the potentially most interesting M-dwarf systems---misaligned or long-period transiting planets---the stellar RME+CLV signals may become quasi-stationary and therefore indistinguishable from the planetary signal, making physically realistic stellar modelling essential for their characterization, even with next generation instruments.


\begin{acknowledgements}
The authors thank the anonymous referee. This work was granted access to the HPC resources of Observatoire de la C\^ote d'Azur, M\'esocentre SIGAMM. TM acknowledges support from the MCIN project PLAtoSOnG (PID2023-146453NB-100).
\end{acknowledgements}

\bibliographystyle{aa}
\bibliography{biblio}

@ARTICLE{Masson2026,
       author = {{Masson}, A. and {Vinatier}, S. and {B{\'e}zard}, B. and {Debras}, F. and {Carmona}, A. and {Lillo-Box}, J. and {Cowan}, N.~B. and {Yariv}, V. and {Allart}, R.},
        title = "{Searching for Molecular Signatures in 14 Transiting Exoplanets with SPIRou}",
      journal = {arXiv e-prints},
         year = 2026,
        month = mar,
          eid = {arXiv:2603.17574},
        pages = {arXiv:2603.17574},
          doi = {10.48550/arXiv.2603.17574},
archivePrefix = {arXiv},
       eprint = {2603.17574},
 primaryClass = {astro-ph.EP},
       adsurl = {https://ui.adsabs.harvard.edu/abs/2026arXiv260317574M}
}

@ARTICLE{Gandhi2020,
       author = {{Gandhi}, Siddharth and {Brogi}, Matteo and {Webb}, Rebecca K.},
        title = "{Seeing above the clouds with high-resolution spectroscopy}",
      journal = {\mnras},
         year = 2020,
        month = oct,
       volume = {498},
       number = {1},
        pages = {194-204},
          doi = {10.1093/mnras/staa2424},
archivePrefix = {arXiv},
       eprint = {2008.11464},
 primaryClass = {astro-ph.EP},
       adsurl = {https://ui.adsabs.harvard.edu/abs/2020MNRAS.498..194G}
}

@ARTICLE{Benneke2019,
       author = {{Benneke}, Bj{\"o}rn and {Knutson}, Heather A. and {Lothringer}, Joshua and {Crossfield}, Ian J.~M. and {Moses}, Julianne I. and {Morley}, Caroline and {Kreidberg}, Laura and {Fulton}, Benjamin J. and {Dragomir}, Diana and {Howard}, Andrew W. and {Wong}, Ian and {D{\'e}sert}, Jean-Michel and {McCullough}, Peter R. and {Kempton}, Eliza M.-R. and {Fortney}, Jonathan and {Gilliland}, Ronald and {Deming}, Drake and {Kammer}, Joshua},
        title = "{A sub-Neptune exoplanet with a low-metallicity methane-depleted atmosphere and Mie-scattering clouds}",
      journal = {Nature Astronomy},
         year = 2019,
        month = jul,
       volume = {3},
        pages = {813-821},
          doi = {10.1038/s41550-019-0800-5},
archivePrefix = {arXiv},
       eprint = {1907.00449},
 primaryClass = {astro-ph.EP},
       adsurl = {https://ui.adsabs.harvard.edu/abs/2019NatAs...3..813B}
}

@ARTICLE{Palle2020,
       author = {{Palle}, E. and {Nortmann}, L. and {Casasayas-Barris}, N. and {Lamp{\'o}n}, M. and {L{\'o}pez-Puertas}, M. and {Caballero}, J.~A. and {Sanz-Forcada}, J. and {Lara}, L.~M. and {Nagel}, E. and {Yan}, F. and {Alonso-Floriano}, F.~J. and {Amado}, P.~J. and {Chen}, G. and {Cifuentes}, C. and {Cort{\'e}s-Contreras}, M. and {Czesla}, S. and {Molaverdikhani}, K. and {Montes}, D. and {Passegger}, V.~M. and {Quirrenbach}, A. and {Reiners}, A. and {Ribas}, I. and {S{\'a}nchez-L{\'o}pez}, A. and {Schweitzer}, A. and {Stangret}, M. and {Zapatero Osorio}, M.~R. and {Zechmeister}, M.},
        title = "{A He I upper atmosphere around the warm Neptune GJ 3470 b}",
      journal = {\aap},
         year = 2020,
        month = jun,
       volume = {638},
          eid = {A61},
        pages = {A61},
          doi = {10.1051/0004-6361/202037719},
archivePrefix = {arXiv},
       eprint = {2004.12812},
 primaryClass = {astro-ph.EP},
       adsurl = {https://ui.adsabs.harvard.edu/abs/2020A&A...638A..61P}
}

@ARTICLE{Stefansson2022,
       author = {{Stef{\`a}nsson}, Gu{\dj}mundur and {Mahadevan}, Suvrath and {Petrovich}, Cristobal and {Winn}, Joshua N. and {Kanodia}, Shubham and {Millholland}, Sarah C. and {Maney}, Marissa and {Ca{\~n}as}, Caleb I. and {Wisniewski}, John and {Robertson}, Paul and {Ninan}, Joe P. and {Ford}, Eric B. and {Bender}, Chad F. and {Blake}, Cullen H. and {Cegla}, Heather and {Cochran}, William D. and {Diddams}, Scott A. and {Dong}, Jiayin and {Endl}, Michael and {Fredrick}, Connor and {Halverson}, Samuel and {Hearty}, Fred and {Hebb}, Leslie and {Hirano}, Teruyuki and {Lin}, Andrea S.~J. and {Logsdon}, Sarah E. and {Lubar}, Emily and {McElwain}, Michael W. and {Metcalf}, Andrew J. and {Monson}, Andrew and {Rajagopal}, Jayadev and {Ramsey}, Lawrence W. and {Roy}, Arpita and {Schwab}, Christian and {Schweiker}, Heidi and {Terrien}, Ryan C. and {Wright}, Jason T.},
        title = "{The Warm Neptune GJ 3470b Has a Polar Orbit}",
      journal = {\apjl},
         year = 2022,
        month = jun,
       volume = {931},
       number = {2},
          eid = {L15},
        pages = {L15},
          doi = {10.3847/2041-8213/ac6e3c},
archivePrefix = {arXiv},
       eprint = {2111.01295},
 primaryClass = {astro-ph.EP},
       adsurl = {https://ui.adsabs.harvard.edu/abs/2022ApJ...931L..15S}
}

@ARTICLE{Bourrier2018,
       author = {{Bourrier}, V. and {Lecavelier des Etangs}, A. and {Ehrenreich}, D. and {Sanz-Forcada}, J. and {Allart}, R. and {Ballester}, G.~E. and {Buchhave}, L.~A. and {Cohen}, O. and {Deming}, D. and {Evans}, T.~M. and {Garc{\'\i}a Mu{\~n}oz}, A. and {Henry}, G.~W. and {Kataria}, T. and {Lavvas}, P. and {Lewis}, N. and {L{\'o}pez-Morales}, M. and {Marley}, M. and {Sing}, D.~K. and {Wakeford}, H.~R.},
        title = "{Hubble PanCET: an extended upper atmosphere of neutral hydrogen around the warm Neptune GJ 3470b}",
      journal = {\aap},
         year = 2018,
        month = dec,
       volume = {620},
          eid = {A147},
        pages = {A147},
          doi = {10.1051/0004-6361/201833675},
archivePrefix = {arXiv},
       eprint = {1812.05119},
 primaryClass = {astro-ph.EP},
       adsurl = {https://ui.adsabs.harvard.edu/abs/2018A&A...620A.147B}
}

@ARTICLE{Bonfils2012,
       author = {{Bonfils}, X. and {Gillon}, M. and {Udry}, S. and {Armstrong}, D. and {Bouchy}, F. and {Delfosse}, X. and {Forveille}, T. and {Fumel}, A. and {Jehin}, E. and {Lendl}, M. and {Lovis}, C. and {Mayor}, M. and {McCormac}, J. and {Neves}, V. and {Pepe}, F. and {Perrier}, C. and {Pollaco}, D. and {Queloz}, D. and {Santos}, N.~C.},
        title = "{A hot Uranus transiting the nearby M dwarf GJ 3470. Detected with HARPS velocimetry. Captured in transit with TRAPPIST photometry}",
      journal = {\aap},
         year = 2012,
        month = oct,
       volume = {546},
          eid = {A27},
        pages = {A27},
          doi = {10.1051/0004-6361/201219623},
archivePrefix = {arXiv},
       eprint = {1206.5307},
 primaryClass = {astro-ph.EP},
       adsurl = {https://ui.adsabs.harvard.edu/abs/2012A&A...546A..27B}
}

@ARTICLE{Souto2017,
 author  = {{Souto}, Diogo and {Cunha}, Katia and {Garc{\'\i}a-Hern{\'a}ndez},
            D.~A. and {Zamora}, O. and {Allende Prieto}, C. and {Smith},
            Verne V. and {Mahadevan}, S. and {Blake}, C. and {Johnson},
            Jennifer A. and {J{\"o}nsson}, H. and {Pinsonneault}, M. and
            {Holtzman}, J. and {Majewski}, S.~R. and {Shetrone}, M. and
            {Teske}, J. and {Nidever}, D. and {Schiavon}, R. and {Sobeck},
            J. and {Fern{\'a}ndez-Trincado}, J.~G. and {Zasowski}, G. and
            {Ivans}, Inese~I.},
 title   = {{Chemical Abundances of M-dwarfs from the APOGEE Survey. I.
            The Exoplanet Hosting Stars Kepler-138 and Kepler-186}},
 journal = {\apj},
 year    = {2017},
 volume  = {835},
 number  = {2},
 pages   = {239},
 doi     = {10.3847/1538-4357/835/2/239}
}

@INPROCEEDINGS{Rainer2018,
       author = {{Rainer}, Monica and {Harutyunyan}, Avet and {Carleo}, Ilaria and {Oliva}, Ernesto and {Benatti}, Serena and {Bignamini}, Andrea and {Claudi}, Riccardo and {Gonzalez-Alvarez}, Esther and {Sanna}, Nicoletta and {Ghedina}, Adriano and {Micela}, Giuseppina and {Molinari}, Emilio and {Tozzi}, Andrea and {Baffa}, Carlo and {Baruffolo}, Andrea and {Buchschacher}, Nicolas and {Cecconi}, Massimo and {Cosentino}, Rosario and {Falcini}, Gilberto and {Fantinel}, Daniela and {Fini}, Luca and {Galli}, Alberto and {Ghinassi}, Francesca and {Giani}, Elisabetta and {Gonzalez}, Carlos and {Gonzalez}, Manuel and {Gratton}, Raffaele and {Guerra}, Jose and {Hernandez Diaz}, Marcos and {Hernandez}, Nauzet and {Iuzzolino}, Marcella and {Lodi}, Marcello and {Malavolta}, Luca and {Maldonado}, Jesus and {Origlia}, Livia and {Perez Ventura}, Hector and {Puglisi}, Alfio and {Riverol}, Carlos and {Riverol}, Luis and {San Juan}, Jose and {Scuderi}, Salvo and {Seeman}, Ulf and {Sozzetti}, Alessandro and {Sozzi}, Mauro},
        title = "{Introducing GOFIO: a DRS for the GIANO-B near-infrared spectrograph}",
    booktitle = {Ground-based and Airborne Instrumentation for Astronomy VII},
         year = 2018,
       editor = {{Evans}, Christopher J. and {Simard}, Luc and {Takami}, Hideki},
       series = {Society of Photo-Optical Instrumentation Engineers (SPIE) Conference Series},
       volume = {10702},
        month = jul,
          eid = {1070266},
        pages = {1070266},
          doi = {10.1117/12.2312130},
       adsurl = {https://ui.adsabs.harvard.edu/abs/2018SPIE10702E..66R}
}

@ARTICLE{2021A&A...650A.201R,
       author = {{Reyl{\'e}}, C. and {Jardine}, K. and {Fouqu{\'e}}, P. and {Caballero}, J.~A. and {Smart}, R.~L. and {Sozzetti}, A.},
        title = "{The 10 parsec sample in the Gaia era}",
      journal = {\aap},
         year = 2021,
        month = jun,
       volume = {650},
          eid = {A201},
        pages = {A201},
          doi = {10.1051/0004-6361/202140985},
archivePrefix = {arXiv},
       eprint = {2104.14972},
 primaryClass = {astro-ph.SR},
       adsurl = {https://ui.adsabs.harvard.edu/abs/2021A&A...650A.201R}
}

@INPROCEEDINGS{Origlia2014,
       author = {{Origlia}, L. and {Oliva}, E. and {Baffa}, C. and {Falcini}, G. and {Giani}, E. and {Massi}, F. and {Montegriffo}, P. and {Sanna}, N. and {Scuderi}, S. and {Sozzi}, M. and {Tozzi}, A. and {Carleo}, I. and {Gratton}, R. and {Ghinassi}, F. and {Lodi}, M.},
        title = "{High resolution near IR spectroscopy with GIANO-TNG}",
    booktitle = {Ground-based and Airborne Instrumentation for Astronomy V},
         year = 2014,
       editor = {{Ramsay}, Suzanne K. and {McLean}, Ian S. and {Takami}, Hideki},
       series = {Society of Photo-Optical Instrumentation Engineers (SPIE) Conference Series},
       volume = {9147},
        month = jul,
          eid = {91471E},
        pages = {91471E},
          doi = {10.1117/12.2054743},
       adsurl = {https://ui.adsabs.harvard.edu/abs/2014SPIE.9147E..1EO}
}

@ARTICLE{Gibson2022,
       author = {{Gibson}, Neale P. and {Nugroho}, Stevanus K. and {Lothringer}, Joshua and {Maguire}, Cathal and {Sing}, David K.},
        title = "{Relative abundance constraints from high-resolution optical transmission spectroscopy of WASP-121b, and a fast model-filtering technique for accelerating retrievals}",
      journal = {\mnras},
         year = 2022,
        month = may,
       volume = {512},
       number = {3},
        pages = {4618-4638},
          doi = {10.1093/mnras/stac091},
archivePrefix = {arXiv},
       eprint = {2201.04025},
 primaryClass = {astro-ph.EP},
       adsurl = {https://ui.adsabs.harvard.edu/abs/2022MNRAS.512.4618G}
}

@ARTICLE{Cheverall2024,
       author = {{Cheverall}, Connor J. and {Madhusudhan}, Nikku},
        title = "{Feasibility of High-resolution Transmission Spectroscopy for Low-velocity Exoplanets}",
      journal = {\aj},
         year = 2024,
        month = jun,
       volume = {167},
       number = {6},
          eid = {272},
        pages = {272},
          doi = {10.3847/1538-3881/ad380c},
archivePrefix = {arXiv},
       eprint = {2403.18894},
 primaryClass = {astro-ph.EP},
       adsurl = {https://ui.adsabs.harvard.edu/abs/2024AJ....167..272C}
}

@ARTICLE{2017A&A...597A..94C,
       author = {{Chiavassa}, A. and {Caldas}, A. and {Selsis}, F. and {Leconte}, J. and {Von Paris}, P. and {Bord{\'e}}, P. and {Magic}, Z. and {Collet}, R. and {Asplund}, M.},
        title = "{Measuring stellar granulation during planet transits}",
      journal = {\aap},
         year = 2017,
        month = jan,
       volume = {597},
          eid = {A94},
        pages = {A94},
          doi = {10.1051/0004-6361/201528018},
archivePrefix = {arXiv},
       eprint = {1609.08966},
 primaryClass = {astro-ph.EP},
       adsurl = {https://ui.adsabs.harvard.edu/abs/2017A&A...597A..94C}
}

@ARTICLE{2010A&A...513A..19F,
       author = {{Freytag}, B. and {Allard}, F. and {Ludwig}, H.-G. and {Homeier}, D. and {Steffen}, M.},
        title = "{The role of convection, overshoot, and gravity waves for the transport of dust in M dwarf and brown dwarf atmospheres}",
      journal = {\aap},
         year = 2010,
        month = apr,
       volume = {513},
          eid = {A19},
        pages = {A19},
          doi = {10.1051/0004-6361/200913354},
archivePrefix = {arXiv},
       eprint = {1002.3437},
 primaryClass = {astro-ph.SR},
       adsurl = {https://ui.adsabs.harvard.edu/abs/2010A&A...513A..19F}
}

@ARTICLE{2002A&A...395...99L,
       author = {{Ludwig}, H.-G. and {Allard}, F. and {Hauschildt}, P.~H.},
        title = "{Numerical simulations of surface convection in a late M-dwarf}",
      journal = {\aap},
         year = 2002,
        month = nov,
       volume = {395},
        pages = {99-115},
          doi = {10.1051/0004-6361:20021153},
archivePrefix = {arXiv},
       eprint = {astro-ph/0208584},
 primaryClass = {astro-ph},
       adsurl = {https://ui.adsabs.harvard.edu/abs/2002A&A...395...99L}
}

@ARTICLE{2009A&A...508.1429W,
       author = {{Wende}, S. and {Reiners}, A. and {Ludwig}, H.-G.},
        title = "{3D simulations of M star atmosphere velocities and their influence on molecular FeH lines}",
      journal = {\aap},
         year = 2009,
        month = dec,
       volume = {508},
       number = {3},
        pages = {1429-1442},
          doi = {10.1051/0004-6361/200913149},
archivePrefix = {arXiv},
       eprint = {0910.3493},
 primaryClass = {astro-ph.SR},
       adsurl = {https://ui.adsabs.harvard.edu/abs/2009A&A...508.1429W}
}

@ARTICLE{2025ApJ...978L..13S,
       author = {{Smitha}, H.~N. and {Shapiro}, Alexander I. and {Witzke}, Veronika and {Kostogryz}, Nadiia M. and {Unruh}, Yvonne C. and {Bhatia}, Tanayveer S. and {Cameron}, Robert and {Seager}, Sara and {Solanki}, Sami K.},
        title = "{First Calculations of Starspot Spectra Based on 3D Radiative Magnetohydrodynamics Simulations}",
      journal = {\apjl},
         year = 2025,
        month = jan,
       volume = {978},
       number = {1},
          eid = {L13},
        pages = {L13},
          doi = {10.3847/2041-8213/ad9aaa},
archivePrefix = {arXiv},
       eprint = {2411.14056},
 primaryClass = {astro-ph.SR},
       adsurl = {https://ui.adsabs.harvard.edu/abs/2025ApJ...978L..13S}
}

@ARTICLE{deKok2013,
       author = {{de Kok}, R.~J. and {Brogi}, M. and {Snellen}, I.~A.~G. and {Birkby}, J. and {Albrecht}, S. and {de Mooij}, E.~J.~W.},
        title = "{Detection of carbon monoxide in the high-resolution day-side spectrum of the exoplanet HD 189733b}",
      journal = {\aap},
         year = 2013,
        month = jun,
       volume = {554},
          eid = {A82},
        pages = {A82},
          doi = {10.1051/0004-6361/201321381},
archivePrefix = {arXiv},
       eprint = {1304.4014},
 primaryClass = {astro-ph.EP},
       adsurl = {https://ui.adsabs.harvard.edu/abs/2013A&A...554A..82D}
}

@ARTICLE{Line2021,
       author = {{Line}, Michael R. and {Brogi}, Matteo and {Bean}, Jacob L. and {Gandhi}, Siddharth and {Zalesky}, Joseph and {Parmentier}, Vivien and {Smith}, Peter and {Mace}, Gregory N. and {Mansfield}, Megan and {Kempton}, Eliza M.-R. and {Fortney}, Jonathan J. and {Shkolnik}, Evgenya and {Patience}, Jennifer and {Rauscher}, Emily and {D{\'e}sert}, Jean-Michel and {Wardenier}, Joost P.},
        title = "{A solar C/O and sub-solar metallicity in a hot Jupiter atmosphere}",
      journal = {\nat},
         year = 2021,
        month = oct,
       volume = {598},
       number = {7882},
        pages = {580-584},
          doi = {10.1038/s41586-021-03912-6},
archivePrefix = {arXiv},
       eprint = {2110.14821},
 primaryClass = {astro-ph.EP},
       adsurl = {https://ui.adsabs.harvard.edu/abs/2021Natur.598..580L}
}

@ARTICLE{Giacobbe2021,
       author = {{Giacobbe}, Paolo and {Brogi}, Matteo and {Gandhi}, Siddharth and {Cubillos}, Patricio E. and {Bonomo}, Aldo S. and {Sozzetti}, Alessandro and {Fossati}, Luca and {Guilluy}, Gloria and {Carleo}, Ilaria and {Rainer}, Monica and {Harutyunyan}, Avet and {Borsa}, Francesco and {Pino}, Lorenzo and {Nascimbeni}, Valerio and {Benatti}, Serena and {Biazzo}, Katia and {Bignamini}, Andrea and {Chubb}, Katy L. and {Claudi}, Riccardo and {Cosentino}, Rosario and {Covino}, Elvira and {Damasso}, Mario and {Desidera}, Silvano and {Fiorenzano}, Aldo F.~M. and {Ghedina}, Adriano and {Lanza}, Antonino F. and {Leto}, Giuseppe and {Maggio}, Antonio and {Malavolta}, Luca and {Maldonado}, Jesus and {Micela}, Giuseppina and {Molinari}, Emilio and {Pagano}, Isabella and {Pedani}, Marco and {Piotto}, Giampaolo and {Poretti}, Ennio and {Scandariato}, Gaetano and {Yurchenko}, Sergei N. and {Fantinel}, Daniela and {Galli}, Alberto and {Lodi}, Marcello and {Sanna}, Nicoletta and {Tozzi}, Andrea},
        title = "{Five carbon- and nitrogen-bearing species in a hot giant planet's atmosphere}",
      journal = {\nat},
         year = 2021,
        month = apr,
       volume = {592},
       number = {7853},
        pages = {205-208},
          doi = {10.1038/s41586-021-03381-x},
archivePrefix = {arXiv},
       eprint = {2104.03352},
 primaryClass = {astro-ph.EP},
       adsurl = {https://ui.adsabs.harvard.edu/abs/2021Natur.592..205G}
}

@ARTICLE{2006MNRAS.368.1087B,
       author = {{Barber}, R.~J. and {Tennyson}, J. and {Harris}, G.~J. and {Tolchenov}, R.~N.},
        title = "{A high-accuracy computed water line list}",
      journal = {\mnras},
         year = 2006,
        month = may,
       volume = {368},
       number = {3},
        pages = {1087-1094},
          doi = {10.1111/j.1365-2966.2006.10184.x},
archivePrefix = {arXiv},
       eprint = {astro-ph/0601236},
 primaryClass = {astro-ph},
       adsurl = {https://ui.adsabs.harvard.edu/abs/2006MNRAS.368.1087B}
}

@ARTICLE{1995stagger,
       author = {{Nordlund}, A. and {Galsgaard}, K.},
        title = "{A 3D MHD code for Parallel Computers}",
      journal = {Tech. rep., Niels Bohr Institute, University of Copenhagen },
         year = 1995,
        month = oct,
       volume = {},
        pages = {},
          doi = {},
archivePrefix = {},
       eprint = {},
 primaryClass = {},
       adsurl = {}
}

@ARTICLE{2018MNRAS.481.1083P,
       author = {{Parsons}, S.~G. and {G{\"a}nsicke}, B.~T. and {Marsh}, T.~R. and {Ashley}, R.~P. and {Breedt}, E. and {Burleigh}, M.~R. and {Copperwheat}, C.~M. and {Dhillon}, V.~S. and {Green}, M.~J. and {Hermes}, J.~J. and {Irawati}, P. and {Kerry}, P. and {Littlefair}, S.~P. and {Rebassa-Mansergas}, A. and {Sahman}, D.~I. and {Schreiber}, M.~R. and {Zorotovic}, M.},
        title = "{The scatter of the M dwarf mass-radius relationship}",
      journal = {\mnras},
         year = 2018,
        month = nov,
       volume = {481},
       number = {1},
        pages = {1083-1096},
          doi = {10.1093/mnras/sty2345},
archivePrefix = {arXiv},
       eprint = {1808.07780},
 primaryClass = {astro-ph.SR},
       adsurl = {https://ui.adsabs.harvard.edu/abs/2018MNRAS.481.1083P}
}

@ARTICLE{2009MmSAI..80..711L,
       author = {{Ludwig}, H.-G. and {Caffau}, E. and {Steffen}, M. and {Freytag}, B. and {Bonifacio}, P. and {Ku{\v{c}}inskas}, A.},
        title = "{The CIFIST 3D model atmosphere grid.}",
      journal = {\memsai},
         year = 2009,
        month = jan,
       volume = {80},
        pages = {711},
          doi = {10.48550/arXiv.0908.4496},
archivePrefix = {arXiv},
       eprint = {0908.4496},
 primaryClass = {astro-ph.SR},
       adsurl = {https://ui.adsabs.harvard.edu/abs/2009MmSAI..80..711L}
}

@ARTICLE{2013A&A...558A..49B,
       author = {{Beeck}, B. and {Cameron}, R.~H. and {Reiners}, A. and {Sch{\"u}ssler}, M.},
        title = "{Three-dimensional simulations of near-surface convection in main-sequence stars. II. Properties of granulation and spectral lines}",
      journal = {\aap},
         year = 2013,
        month = oct,
       volume = {558},
          eid = {A49},
        pages = {A49},
          doi = {10.1051/0004-6361/201321345},
archivePrefix = {arXiv},
       eprint = {1308.4873},
 primaryClass = {astro-ph.SR},
       adsurl = {https://ui.adsabs.harvard.edu/abs/2013A&A...558A..49B}
}

@ARTICLE{2013A&A...559A.104T,
       author = {{Tremblay}, P.-E. and {Ludwig}, H.-G. and {Steffen}, M. and {Freytag}, B.},
        title = "{Spectroscopic analysis of DA white dwarfs with 3D model atmospheres}",
      journal = {\aap},
         year = 2013,
        month = nov,
       volume = {559},
          eid = {A104},
        pages = {A104},
          doi = {10.1051/0004-6361/201322318},
archivePrefix = {arXiv},
       eprint = {1309.0886},
 primaryClass = {astro-ph.SR},
       adsurl = {https://ui.adsabs.harvard.edu/abs/2013A&A...559A.104T}
}

@ARTICLE{Gandhi2019,
       author = {{Gandhi}, Siddharth and {Madhusudhan}, Nikku},
        title = "{New avenues for thermal inversions in atmospheres of hot Jupiters}",
      journal = {\mnras},
         year = 2019,
        month = jun,
       volume = {485},
       number = {4},
        pages = {5817-5830},
          doi = {10.1093/mnras/stz751},
archivePrefix = {arXiv},
       eprint = {1903.11603},
 primaryClass = {astro-ph.EP},
       adsurl = {https://ui.adsabs.harvard.edu/abs/2019MNRAS.485.5817G}
}

@ARTICLE{2024A&A...688A..52E,
       author = {{Eitner}, Philipp and {Bergemann}, Maria and {Hoppe}, Richard and {Nordlund}, {\r{A}}ke and {Plez}, Bertrand and {Klevas}, Jonas},
        title = "{M3DIS - A grid of 3D radiation-hydrodynamics stellar atmosphere models for stellar surveys. I. Procedure, validation, and the Sun}",
      journal = {\aap},
         year = 2024,
        month = aug,
       volume = {688},
          eid = {A52},
        pages = {A52},
          doi = {10.1051/0004-6361/202348448},
archivePrefix = {arXiv},
       eprint = {2405.06338},
 primaryClass = {astro-ph.SR},
       adsurl = {https://ui.adsabs.harvard.edu/abs/2024A&A...688A..52E}
}

@ARTICLE{2024A&A...688A.212R,
       author = {{Rodr{\'\i}guez D{\'\i}az}, Luisa F. and {Lagae}, Cis and {Amarsi}, Anish M. and {Bigot}, Lionel and {Zhou}, Yixiao and {Aguirre B{\o}rsen-Koch}, V{\'\i}ctor and {Lind}, Karin and {Trampedach}, Regner and {Collet}, Remo},
        title = "{An extended and refined grid of 3D STAGGER model atmospheres. Processed snapshots for stellar spectroscopy}",
      journal = {\aap},
         year = 2024,
        month = aug,
       volume = {688},
          eid = {A212},
        pages = {A212},
          doi = {10.1051/0004-6361/202348480},
archivePrefix = {arXiv},
       eprint = {2405.07872},
 primaryClass = {astro-ph.SR},
       adsurl = {https://ui.adsabs.harvard.edu/abs/2024A&A...688A.212R}
}

@INPROCEEDINGS{2003ASPC..288..519S,
       author = {{Stein}, R.~F. and {Nordlund}, {\r{A}}.},
        title = "{Radiative Transfer in 3D Numerical Simulations}",
    booktitle = {Stellar Atmosphere Modeling},
         year = 2003,
       editor = {{Hubeny}, Ivan and {Mihalas}, Dimitri and {Werner}, Klaus},
       series = {Astronomical Society of the Pacific Conference Series},
       volume = {288},
        month = jan,
        pages = {519},
          doi = {10.48550/arXiv.astro-ph/0209510},
archivePrefix = {arXiv},
       eprint = {astro-ph/0209510},
 primaryClass = {astro-ph},
       adsurl = {https://ui.adsabs.harvard.edu/abs/2003ASPC..288..519S}
}

@ARTICLE{1982A&A...107....1N,
       author = {{Nordlund}, A.},
        title = "{Numerical simulations of the solar granulation. I. Basic equations and methods.}",
      journal = {\aap},
         year = 1982,
        month = mar,
       volume = {107},
        pages = {1-10},
       adsurl = {https://ui.adsabs.harvard.edu/abs/1982A&A...107....1N}
}

@ARTICLE{2000ApJ...536..465S,
       author = {{Skartlien}, R.},
        title = "{A Multigroup Method for Radiation with Scattering in Three-Dimensional Hydrodynamic Simulations}",
      journal = {\apj},
         year = 2000,
        month = jun,
       volume = {536},
       number = {1},
        pages = {465-480},
          doi = {10.1086/308934},
       adsurl = {https://ui.adsabs.harvard.edu/abs/2000ApJ...536..465S}
}

@ARTICLE{2010A&A...517A..49H,
       author = {{Hayek}, W. and {Asplund}, M. and {Carlsson}, M. and {Trampedach}, R. and {Collet}, R. and {Gudiksen}, B.~V. and {Hansteen}, V.~H. and {Leenaarts}, J.},
        title = "{Radiative transfer with scattering for domain-decomposed 3D MHD simulations of cool stellar atmospheres. Numerical methods and application to the quiet, non-magnetic, surface of a solar-type star}",
      journal = {\aap},
         year = 2010,
        month = jul,
       volume = {517},
          eid = {A49},
        pages = {A49},
          doi = {10.1051/0004-6361/201014210},
archivePrefix = {arXiv},
       eprint = {1007.2760},
 primaryClass = {astro-ph.SR},
       adsurl = {https://ui.adsabs.harvard.edu/abs/2010A&A...517A..49H}
}

@ARTICLE{2008A&A...486..951G,
       author = {{Gustafsson}, B. and {Edvardsson}, B. and {Eriksson}, K. and {J{\o}rgensen}, U.~G. and {Nordlund}, {\r{A}}. and {Plez}, B.},
        title = "{A grid of MARCS model atmospheres for late-type stars. I. Methods and general properties}",
      journal = {\aap},
         year = 2008,
        month = aug,
       volume = {486},
       number = {3},
        pages = {951-970},
          doi = {10.1051/0004-6361:200809724},
archivePrefix = {arXiv},
       eprint = {0805.0554},
 primaryClass = {astro-ph},
       adsurl = {https://ui.adsabs.harvard.edu/abs/2008A&A...486..951G}
}

@ARTICLE{2009ARA&A..47..481A,
       author = {{Asplund}, Martin and {Grevesse}, Nicolas and {Sauval}, A. Jacques and {Scott}, Pat},
        title = "{The Chemical Composition of the Sun}",
      journal = {\araa},
         year = 2009,
        month = sep,
       volume = {47},
       number = {1},
        pages = {481-522},
          doi = {10.1146/annurev.astro.46.060407.145222},
archivePrefix = {arXiv},
       eprint = {0909.0948},
 primaryClass = {astro-ph.SR},
       adsurl = {https://ui.adsabs.harvard.edu/abs/2009ARA&A..47..481A}
}

@ARTICLE{2018MNRAS.475.3369C,
       author = {{Collet}, R. and {Nordlund}, {\r{A}}. and {Asplund}, M. and {Hayek}, W. and {Trampedach}, R.},
        title = "{The benchmark halo giant HD 122563: CNO abundances revisited with three-dimensional hydrodynamic model stellar atmospheres}",
      journal = {\mnras},
         year = 2018,
        month = apr,
       volume = {475},
       number = {3},
        pages = {3369-3392},
          doi = {10.1093/mnras/sty002},
archivePrefix = {arXiv},
       eprint = {1712.08099},
 primaryClass = {astro-ph.SR},
       adsurl = {https://ui.adsabs.harvard.edu/abs/2018MNRAS.475.3369C}
}

@ARTICLE{2013A&A...557A..26M,
       author = {{Magic}, Z. and {Collet}, R. and {Asplund}, M. and {Trampedach}, R. and {Hayek}, W. and {Chiavassa}, A. and {Stein}, R.~F. and {Nordlund}, {\r{A}}.},
        title = "{The Stagger-grid: A grid of 3D stellar atmosphere models. I. Methods and general properties}",
      journal = {\aap},
         year = 2013,
        month = sep,
       volume = {557},
          eid = {A26},
        pages = {A26},
          doi = {10.1051/0004-6361/201321274},
archivePrefix = {arXiv},
       eprint = {1302.2621},
 primaryClass = {astro-ph.SR},
       adsurl = {https://ui.adsabs.harvard.edu/abs/2013A&A...557A..26M}
}

@ARTICLE{2024ApJ...970...24S,
       author = {{Stein}, Robert F. and {Nordlund}, {\r{A}}ke and {Collet}, Remo and {Trampedach}, Regner},
        title = "{The Stagger Code for Accurate and Efficient, Radiation-coupled Magnetohydrodynamic Simulations}",
      journal = {\apj},
         year = 2024,
        month = jul,
       volume = {970},
       number = {1},
          eid = {24},
        pages = {24},
          doi = {10.3847/1538-4357/ad4706},
archivePrefix = {arXiv},
       eprint = {2405.02483},
 primaryClass = {astro-ph.IM},
       adsurl = {https://ui.adsabs.harvard.edu/abs/2024ApJ...970...24S}
}

@ARTICLE{2018MNRAS.480.2597P,
       author = {{Polyansky}, Oleg L. and {Kyuberis}, Aleksandra A. and {Zobov}, Nikolai F. and {Tennyson}, Jonathan and {Yurchenko}, Sergei N. and {Lodi}, Lorenzo},
        title = "{ExoMol molecular line lists XXX: a complete high-accuracy line list for water}",
      journal = {\mnras},
         year = 2018,
        month = oct,
       volume = {480},
       number = {2},
        pages = {2597-2608},
          doi = {10.1093/mnras/sty1877},
archivePrefix = {arXiv},
       eprint = {1807.04529},
 primaryClass = {astro-ph.EP},
       adsurl = {https://ui.adsabs.harvard.edu/abs/2018MNRAS.480.2597P}
}

@misc{2016ascl.soft05004M,
       author = {{Masseron}, Thomas and {Merle}, Thibault and {Hawkins}, Keith},
        title = "{BACCHUS: Brussels Automatic Code for Characterizing High accUracy Spectra}",
 howpublished = {Astrophysics Source Code Library, record ascl:1605.004},
         year = 2016,
        month = may,
          eid = {ascl:1605.004},
archivePrefix = {ascl},
       eprint = {1605.004},
       adsurl = {https://ui.adsabs.harvard.edu/abs/2016ascl.soft05004M}
}

@ARTICLE{2018A&A...620A.180R,
       author = {{Rajpurohit}, A.~S. and {Allard}, F. and {Rajpurohit}, S. and {Sharma}, R. and {Teixeira}, G.~D.~C. and {Mousis}, O. and {Rajpurohit}, Kamlesh},
        title = "{Exploring the stellar properties of M dwarfs with high-resolution spectroscopy from the optical to the near-infrared}",
      journal = {\aap},
         year = 2018,
        month = dec,
       volume = {620},
          eid = {A180},
        pages = {A180},
          doi = {10.1051/0004-6361/201833500},
archivePrefix = {arXiv},
       eprint = {1810.13252},
 primaryClass = {astro-ph.SR},
       adsurl = {https://ui.adsabs.harvard.edu/abs/2018A&A...620A.180R}
}

@ARTICLE{2010JQSRT.111.2139R,
       author = {{Rothman}, L.~S. and {Gordon}, I.~E. and {Barber}, R.~J. and {Dothe}, H. and {Gamache}, R.~R. and {Goldman}, A. and {Perevalov}, V.~I. and {Tashkun}, S.~A. and {Tennyson}, J.},
        title = "{HITEMP, the high-temperature molecular spectroscopic database}",
      journal = {\jqsrt},
         year = 2010,
        month = oct,
       volume = {111},
        pages = {2139-2150},
          doi = {10.1016/j.jqsrt.2010.05.001},
       adsurl = {https://ui.adsabs.harvard.edu/abs/2010JQSRT.111.2139R}
}

@ARTICLE{2000A&A...359..755A,
       author = {{Asplund}, M.},
        title = "{Line formation in solar granulation. III. The photospheric Si and meteoritic Fe abundances}",
      journal = {\aap},
         year = 2000,
        month = jul,
       volume = {359},
        pages = {755-758},
          doi = {10.48550/arXiv.astro-ph/0005322},
archivePrefix = {arXiv},
       eprint = {astro-ph/0005322},
 primaryClass = {astro-ph},
       adsurl = {https://ui.adsabs.harvard.edu/abs/2000A&A...359..755A}
}

@ARTICLE{2009A&A...506.1351C,
       author = {{Chiavassa}, A. and {Plez}, B. and {Josselin}, E. and {Freytag}, B.},
        title = "{Radiative hydrodynamics simulations of red supergiant stars. I. interpretation of interferometric observations}",
      journal = {\aap},
         year = 2009,
        month = nov,
       volume = {506},
       number = {3},
        pages = {1351-1365},
          doi = {10.1051/0004-6361/200911780},
archivePrefix = {arXiv},
       eprint = {0907.1860},
 primaryClass = {astro-ph.SR},
       adsurl = {https://ui.adsabs.harvard.edu/abs/2009A&A...506.1351C}
}

@ARTICLE{Smith2021,
       author = {{Smith}, Verne V. and {Bizyaev}, Dmitry and {Cunha}, Katia and {Shetrone}, Matthew D. and {Souto}, Diogo and {Allende Prieto}, Carlos and {Masseron}, Thomas and {M{\'e}sz{\'a}ros}, Szabolcs and {J{\"o}nsson}, Henrik and {Hasselquist}, Sten and {Osorio}, Yeisson and {Garc{\'\i}a-Hern{\'a}ndez}, D.~A. and {Plez}, Bertrand and {Beaton}, Rachael L. and {Holtzman}, Jon and {Majewski}, Steven R. and {Stringfellow}, Guy S. and {Sobeck}, Jennifer},
        title = "{The APOGEE Data Release 16 Spectral Line List}",
      journal = {\aj},
         year = 2021,
        month = jun,
       volume = {161},
       number = {6},
          eid = {254},
        pages = {254},
          doi = {10.3847/1538-3881/abefdc},
archivePrefix = {arXiv},
       eprint = {2103.10112},
 primaryClass = {astro-ph.SR},
       adsurl = {https://ui.adsabs.harvard.edu/abs/2021AJ....161..254S}
}

@ARTICLE{Heiter2021,
       author = {{Heiter}, U. and {Lind}, K. and {Bergemann}, M. and {Asplund}, M. and {Mikolaitis}, {\v{S}}. and {Barklem}, P.~S. and {Masseron}, T. and {de Laverny}, P. and {Magrini}, L. and {Edvardsson}, B. and {J{\"o}nsson}, H. and {Pickering}, J.~C. and {Ryde}, N. and {Bayo Ar{\'a}n}, A. and {Bensby}, T. and {Casey}, A.~R. and {Feltzing}, S. and {Jofr{\'e}}, P. and {Korn}, A.~J. and {Pancino}, E. and {Damiani}, F. and {Lanzafame}, A. and {Lardo}, C. and {Monaco}, L. and {Morbidelli}, L. and {Smiljanic}, R. and {Worley}, C. and {Zaggia}, S. and {Randich}, S. and {Gilmore}, G.~F.},
        title = "{Atomic data for the Gaia-ESO Survey}",
      journal = {\aap},
         year = 2021,
        month = jan,
       volume = {645},
          eid = {A106},
        pages = {A106},
          doi = {10.1051/0004-6361/201936291},
archivePrefix = {arXiv},
       eprint = {2011.02049},
 primaryClass = {astro-ph.IM},
       adsurl = {https://ui.adsabs.harvard.edu/abs/2021A&A...645A.106H}
}

@ARTICLE{Yurchenko2018_C2,
       author = {{Yurchenko}, Sergei N. and {Szab{\'o}}, Istv{\'a}n and {Pyatenko}, Elizaveta and {Tennyson}, Jonathan},
        title = "{ExoMol line lists XXXI: spectroscopy of lowest eights electronic states of C$_{2}$}",
      journal = {\mnras},
         year = 2018,
        month = nov,
       volume = {480},
       number = {3},
        pages = {3397-3411},
          doi = {10.1093/mnras/sty2050},
archivePrefix = {arXiv},
       eprint = {1812.07116},
 primaryClass = {astro-ph.SR},
       adsurl = {https://ui.adsabs.harvard.edu/abs/2018MNRAS.480.3397Y}
}

@ARTICLE{Brooke2016,
       author = {{Brooke}, James S.~A. and {Bernath}, Peter F. and {Western}, Colin M. and {Sneden}, Christopher and {Af{\c{s}}ar}, Melike and {Li}, Gang and {Gordon}, Iouli E.},
        title = "{Line strengths of rovibrational and rotational transitions in the X$^{2}$ {\ensuremath{\Pi}} ground state of OH}",
      journal = {\jqsrt},
         year = 2016,
        month = jan,
       volume = {168},
        pages = {142-157},
          doi = {10.1016/j.jqsrt.2015.07.021},
       adsurl = {https://ui.adsabs.harvard.edu/abs/2016JQSRT.168..142B}
}

@ARTICLE{Yurchenko2018_SiH,
       author = {{Yurchenko}, Sergei N. and {Sinden}, Frances and {Lodi}, Lorenzo and {Hill}, Christian and {Gorman}, Maire N. and {Tennyson}, Jonathan},
        title = "{ExoMol line lists XXIV: a new hot line list for silicon monohydride, SiH}",
      journal = {\mnras},
         year = 2018,
        month = feb,
       volume = {473},
       number = {4},
        pages = {5324-5333},
          doi = {10.1093/mnras/stx2738},
archivePrefix = {arXiv},
       eprint = {1710.06964},
 primaryClass = {astro-ph.SR},
       adsurl = {https://ui.adsabs.harvard.edu/abs/2018MNRAS.473.5324Y}
}

@ARTICLE{Burrows2002,
       author = {{Burrows}, Adam and {Ram}, R.~S. and {Bernath}, Peter and {Sharp}, C.~M. and {Milsom}, J.~A.},
        title = "{New CrH Opacities for the Study of L and Brown Dwarf Atmospheres}",
      journal = {\apj},
         year = 2002,
        month = oct,
       volume = {577},
       number = {2},
        pages = {986-992},
          doi = {10.1086/342242},
archivePrefix = {arXiv},
       eprint = {astro-ph/0206159},
 primaryClass = {astro-ph},
       adsurl = {https://ui.adsabs.harvard.edu/abs/2002ApJ...577..986B}
}

@ARTICLE{Sneden2014,
       author = {{Sneden}, Christopher and {Lucatello}, Sara and {Ram}, Ram S. and {Brooke}, James S.~A. and {Bernath}, Peter},
        title = "{Line Lists for the A $^{2}${\ensuremath{\Pi}}-X $^{2}${\ensuremath{\Sigma}}$^{+}$ (Red) and B $^{2}${\ensuremath{\Sigma}}$^{+}$-X $^{2}${\ensuremath{\Sigma}}$^{+}$ (Violet) Systems of CN, $^{13}$C$^{14}$N, and $^{12}$C$^{15}$N, and Application to Astronomical Spectra}",
      journal = {\apjs},
         year = 2014,
        month = oct,
       volume = {214},
       number = {2},
          eid = {26},
        pages = {26},
          doi = {10.1088/0067-0049/214/2/26},
archivePrefix = {arXiv},
       eprint = {1408.3828},
 primaryClass = {astro-ph.SR},
       adsurl = {https://ui.adsabs.harvard.edu/abs/2014ApJS..214...26S}
}

@ARTICLE{Barton2013,
       author = {{Barton}, Emma J. and {Yurchenko}, Sergei N. and {Tennyson}, Jonathan},
        title = "{ExoMol line lists - II. The ro-vibrational spectrum of SiO}",
      journal = {\mnras},
         year = 2013,
        month = sep,
       volume = {434},
       number = {2},
        pages = {1469-1475},
          doi = {10.1093/mnras/stt1105},
archivePrefix = {arXiv},
       eprint = {1307.2300},
 primaryClass = {astro-ph.SR},
       adsurl = {https://ui.adsabs.harvard.edu/abs/2013MNRAS.434.1469B}
}

@ARTICLE{Dulick2003,
       author = {{Dulick}, M. and {Bauschlicher}, Jr., C.~W. and {Burrows}, Adam and {Sharp}, C.~M. and {Ram}, R.~S. and {Bernath}, Peter},
        title = "{Line Intensities and Molecular Opacities of the FeH F $^{4}${\ensuremath{\Delta}}$_{i}$-X $^{4}${\ensuremath{\Delta}}$_{i}$ Transition}",
      journal = {\apj},
         year = 2003,
        month = sep,
       volume = {594},
       number = {2},
        pages = {651-663},
          doi = {10.1086/376791},
archivePrefix = {arXiv},
       eprint = {astro-ph/0305162},
 primaryClass = {astro-ph},
       adsurl = {https://ui.adsabs.harvard.edu/abs/2003ApJ...594..651D}
}

@ARTICLE{Hargreaves2010,
       author = {{Hargreaves}, Robert J. and {Hinkle}, Kenneth H. and {Bauschlicher}, Jr., Charles W. and {Wende}, Sebastian and {Seifahrt}, Andreas and {Bernath}, Peter F.},
        title = "{High-resolution 1.6 {\ensuremath{\mu}}m Spectra of FeH in M and L Dwarfs}",
      journal = {\aj},
         year = 2010,
        month = oct,
       volume = {140},
       number = {4},
        pages = {919-924},
          doi = {10.1088/0004-6256/140/4/919},
       adsurl = {https://ui.adsabs.harvard.edu/abs/2010AJ....140..919H}
}

@ARTICLE{Yadin2012,
       author = {{Yadin}, Benjamin and {Veness}, Thomas and {Conti}, Pierandrea and {Hill}, Christian and {Yurchenko}, Sergei N. and {Tennyson}, Jonathan},
        title = "{ExoMol line lists - I. The rovibrational spectrum of BeH, MgH and CaH in the X $^{2}${\ensuremath{\Sigma}}$^{+}$ state}",
      journal = {\mnras},
         year = 2012,
        month = sep,
       volume = {425},
       number = {1},
        pages = {34-43},
          doi = {10.1111/j.1365-2966.2012.21367.x},
archivePrefix = {arXiv},
       eprint = {1204.0137},
 primaryClass = {astro-ph.SR},
       adsurl = {https://ui.adsabs.harvard.edu/abs/2012MNRAS.425...34Y}
}

@ARTICLE{McKemmish2019,
       author = {{McKemmish}, Laura K. and {Masseron}, Thomas and {Hoeijmakers}, H. Jens and {P{\'e}rez-Mesa}, V{\'\i}ctor and {Grimm}, Simon L. and {Yurchenko}, Sergei N. and {Tennyson}, Jonathan},
        title = "{ExoMol molecular line lists - XXXIII. The spectrum of Titanium Oxide}",
      journal = {\mnras},
         year = 2019,
        month = sep,
       volume = {488},
       number = {2},
        pages = {2836-2854},
          doi = {10.1093/mnras/stz1818},
archivePrefix = {arXiv},
       eprint = {1905.04587},
 primaryClass = {astro-ph.SR},
       adsurl = {https://ui.adsabs.harvard.edu/abs/2019MNRAS.488.2836M}
}

@ARTICLE{Yurchenko2018_AlH,
       author = {{Yurchenko}, Sergei N. and {Williams}, Henry and {Leyland}, Paul C. and {Lodi}, Lorenzo and {Tennyson}, Jonathan},
        title = "{ExoMol linelists XXVIII: the rovibronic spectrum of AlH}",
      journal = {\mnras},
         year = 2018,
        month = sep,
       volume = {479},
       number = {1},
        pages = {1401-1411},
          doi = {10.1093/mnras/sty1524},
archivePrefix = {arXiv},
       eprint = {1806.03279},
 primaryClass = {astro-ph.SR},
       adsurl = {https://ui.adsabs.harvard.edu/abs/2018MNRAS.479.1401Y}
}

@ARTICLE{McKemmish2016,
       author = {{McKemmish}, Laura K. and {Yurchenko}, Sergei N. and {Tennyson}, Jonathan},
        title = "{ExoMol line lists - XVIII. The high-temperature spectrum of VO}",
      journal = {\mnras},
         year = 2016,
        month = nov,
       volume = {463},
       number = {1},
        pages = {771-793},
          doi = {10.1093/mnras/stw1969},
archivePrefix = {arXiv},
       eprint = {1609.06120},
 primaryClass = {astro-ph.SR},
       adsurl = {https://ui.adsabs.harvard.edu/abs/2016MNRAS.463..771M}
}

@ARTICLE{Masseron2014,
       author = {{Masseron}, T. and {Plez}, B. and {Van Eck}, S. and {Colin}, R. and {Daoutidis}, I. and {Godefroid}, M. and {Coheur}, P.-F. and {Bernath}, P. and {Jorissen}, A. and {Christlieb}, N.},
        title = "{CH in stellar atmospheres: an extensive linelist}",
      journal = {\aap},
         year = 2014,
        month = nov,
       volume = {571},
          eid = {A47},
        pages = {A47},
          doi = {10.1051/0004-6361/201423956},
archivePrefix = {arXiv},
       eprint = {1410.4005},
 primaryClass = {astro-ph.SR},
       adsurl = {https://ui.adsabs.harvard.edu/abs/2014A&A...571A..47M}
}

@ARTICLE{2012MNRAS.425...34Y,
       author = {{Yadin}, Benjamin and {Veness}, Thomas and {Conti}, Pierandrea and {Hill}, Christian and {Yurchenko}, Sergei N. and {Tennyson}, Jonathan},
        title = "{ExoMol line lists - I. The rovibrational spectrum of BeH, MgH and CaH in the X $^{2}${\ensuremath{\Sigma}}$^{+}$ state}",
      journal = {\mnras},
         year = 2012,
        month = sep,
       volume = {425},
       number = {1},
        pages = {34-43},
          doi = {10.1111/j.1365-2966.2012.21367.x},
archivePrefix = {arXiv},
       eprint = {1204.0137},
 primaryClass = {astro-ph.SR},
       adsurl = {https://ui.adsabs.harvard.edu/abs/2012MNRAS.425...34Y}
}

@ARTICLE{Yousefi2018,
       author = {{Yousefi}, Mahdi and {Bernath}, Peter F. and {Hodges}, James and {Masseron}, Thomas},
        title = "{A new line list for the A$^{2}${\ensuremath{\Sigma}}$^{+}$ -X$^{2}$ {\ensuremath{\Pi}} electronic transition of OH}",
      journal = {\jqsrt},
         year = 2018,
        month = sep,
       volume = {217},
        pages = {416-424},
          doi = {10.1016/j.jqsrt.2018.06.016},
       adsurl = {https://ui.adsabs.harvard.edu/abs/2018JQSRT.217..416Y}
}

@ARTICLE{Shayesteh2013,
       author = {{Shayesteh}, Alireza and {Ram}, Ram S. and {Bernath}, Peter F.},
        title = "{Fourier transform emission spectra of the A$^{2}${\ensuremath{\Pi}} {\textrightarrow} X$^{2}${\ensuremath{\Sigma}}$^{+}$ and B$^{2}${\ensuremath{\Sigma}}$^{+}$ {\textrightarrow} X$^{2}${\ensuremath{\Sigma}}$^{+}$ band systems of CaH}",
      journal = {Journal of Molecular Spectroscopy},
         year = 2013,
        month = jun,
       volume = {288},
        pages = {46-51},
          doi = {10.1016/j.jms.2013.04.009},
       adsurl = {https://ui.adsabs.harvard.edu/abs/2013JMoSp.288...46S}
}

@ARTICLE{Hinkle2013,
       author = {{Hinkle}, Kenneth H. and {Wallace}, Lloyd and {Ram}, Ram S. and {Bernath}, Peter F. and {Sneden}, Christopher and {Lucatello}, Sara},
        title = "{The Magnesium Isotopologues of MgH in the A $^{2}${\ensuremath{\Pi}}-X $^{2}${\ensuremath{\Sigma}}$^{+}$ System}",
      journal = {\apjs},
         year = 2013,
        month = aug,
       volume = {207},
       number = {2},
          eid = {26},
        pages = {26},
          doi = {10.1088/0067-0049/207/2/26},
       adsurl = {https://ui.adsabs.harvard.edu/abs/2013ApJS..207...26H}
}

@misc{2012ascl.soft05004P,
       author = {{Plez}, B.},
        title = "{Turbospectrum: Code for spectral synthesis}",
 howpublished = {Astrophysics Source Code Library, record ascl:1205.004},
         year = 2012,
        month = may,
          eid = {ascl:1205.004},
archivePrefix = {ascl},
       eprint = {1205.004},
       adsurl = {https://ui.adsabs.harvard.edu/abs/2012ascl.soft05004P}
}

@ARTICLE{2010A&A...524A..93C,
       author = {{Chiavassa}, A. and {Collet}, R. and {Casagrande}, L. and {Asplund}, M.},
        title = "{Three-dimensional hydrodynamical simulations of red giant stars: semi-global models for interpreting interferometric observations}",
      journal = {\aap},
         year = 2010,
        month = dec,
       volume = {524},
          eid = {A93},
        pages = {A93},
          doi = {10.1051/0004-6361/201015507},
archivePrefix = {arXiv},
       eprint = {1009.1745},
 primaryClass = {astro-ph.SR},
       adsurl = {https://ui.adsabs.harvard.edu/abs/2010A&A...524A..93C}
}

@ARTICLE{2012JQSRT.113..929F,
       author = {{Furtenbacher}, Tibor and {Cs{\'a}sz{\'a}r}, Attila G.},
        title = "{MARVEL: Measured active rotational-vibrational energy levels. II. Algorithmic improvements}",
      journal = {\jqsrt},
         year = 2012,
        month = jul,
       volume = {113},
       number = {11},
        pages = {929-935},
          doi = {10.1016/j.jqsrt.2012.01.005},
       adsurl = {https://ui.adsabs.harvard.edu/abs/2012JQSRT.113..929F}
}

@ARTICLE{2009LRSP....6....2N,
       author = {{Nordlund}, {\r{A}}ke and {Stein}, Robert F. and {Asplund}, Martin},
        title = "{Solar Surface Convection}",
      journal = {Living Reviews in Solar Physics},
         year = 2009,
        month = dec,
       volume = {6},
       number = {1},
          eid = {2},
        pages = {2},
          doi = {10.12942/lrsp-2009-2},
       adsurl = {https://ui.adsabs.harvard.edu/abs/2009LRSP....6....2N}
}

@ARTICLE{2015MNRAS.449.3613P,
       author = {{Patrascu}, Andrei T. and {Yurchenko}, Sergei N. and {Tennyson}, Jonathan},
        title = "{ExoMol molecular line lists - IX. The spectrum of AlO}",
      journal = {\mnras},
         year = 2015,
        month = jun,
       volume = {449},
       number = {4},
        pages = {3613-3619},
          doi = {10.1093/mnras/stv507},
archivePrefix = {arXiv},
       eprint = {1504.02938},
 primaryClass = {astro-ph.GA},
       adsurl = {https://ui.adsabs.harvard.edu/abs/2015MNRAS.449.3613P}
}

@ARTICLE{2018JQSRT.217...29F,
       author = {{Fernando}, Anton M. and {Bernath}, Peter F. and {Hodges}, James N. and {Masseron}, Thomas},
        title = "{A new linelist for the A$^{3}${\ensuremath{\Pi}}-X$^{3}${\ensuremath{\Sigma}}$^{-}$ transition of the NH free radical}",
      journal = {\jqsrt},
         year = 2018,
        month = sep,
       volume = {217},
        pages = {29-34},
          doi = {10.1016/j.jqsrt.2018.05.021},
       adsurl = {https://ui.adsabs.harvard.edu/abs/2018JQSRT.217...29F}
}

@ARTICLE{2022ApJ...933...99B,
       author = {{Bernath}, P.~F. and {Dodangodage}, R. and {Li{\'e}vin}, J.},
        title = "{S-type Stars: LaO Line List for the B$^{2}${\ensuremath{\Sigma}}$^{+}$-X$^{2}${\ensuremath{\Sigma}}$^{+}$ Band System}",
      journal = {\apj},
         year = 2022,
        month = jul,
       volume = {933},
       number = {1},
          eid = {99},
        pages = {99},
          doi = {10.3847/1538-4357/ac731f},
       adsurl = {https://ui.adsabs.harvard.edu/abs/2022ApJ...933...99B}
}

@ARTICLE{2023ApJ...953..181B,
       author = {{Bernath}, P.~F. and {Dodangodage}, R. and {Li{\'e}vin}, J.},
        title = "{S-type Stars: Line List for the A$^{2}${\ensuremath{\Pi}}-X$^{2}${\ensuremath{\Sigma}}$^{+}$ Band System of LaO}",
      journal = {\apj},
         year = 2023,
        month = aug,
       volume = {953},
       number = {2},
          eid = {181},
        pages = {181},
          doi = {10.3847/1538-4357/ace68a},
       adsurl = {https://ui.adsabs.harvard.edu/abs/2023ApJ...953..181B}
}

@ARTICLE{2006AJ....132.2360H,
       author = {{Henry}, Todd J. and {Jao}, Wei-Chun and {Subasavage}, John P. and {Beaulieu}, Thomas D. and {Ianna}, Philip A. and {Costa}, Edgardo and {M{\'e}ndez}, Ren{\'e} A.},
        title = "{The Solar Neighborhood. XVII. Parallax Results from the CTIOPI 0.9 m Program: 20 New Members of the RECONS 10 Parsec Sample}",
      journal = {\aj},
         year = 2006,
        month = dec,
       volume = {132},
       number = {6},
        pages = {2360-2371},
          doi = {10.1086/508233},
archivePrefix = {arXiv},
       eprint = {astro-ph/0608230},
 primaryClass = {astro-ph},
       adsurl = {https://ui.adsabs.harvard.edu/abs/2006AJ....132.2360H}
}

@ARTICLE{2015ApJ...807...45D,
       author = {{Dressing}, Courtney D. and {Charbonneau}, David},
        title = "{The Occurrence of Potentially Habitable Planets Orbiting M Dwarfs Estimated from the Full Kepler Dataset and an Empirical Measurement of the Detection Sensitivity}",
      journal = {\apj},
         year = 2015,
        month = jul,
       volume = {807},
       number = {1},
          eid = {45},
        pages = {45},
          doi = {10.1088/0004-637X/807/1/45},
archivePrefix = {arXiv},
       eprint = {1501.01623},
 primaryClass = {astro-ph.EP},
       adsurl = {https://ui.adsabs.harvard.edu/abs/2015ApJ...807...45D}
}

@ARTICLE{2023A&A...670A..84K,
       author = {{Kossakowski}, D. and {K{\"u}rster}, M. and {Trifonov}, T. and {Henning}, Th. and {Kemmer}, J. and {Caballero}, J.~A. and {Burn}, R. and {Sabotta}, S. and {Crouse}, J.~S. and {Fauchez}, T.~J. and {Nagel}, E. and {Kaminski}, A. and {Herrero}, E. and {Rodr{\'\i}guez}, E. and {Gonz{\'a}lez-{\'A}lvarez}, E. and {Quirrenbach}, A. and {Amado}, P.~J. and {Ribas}, I. and {Reiners}, A. and {Aceituno}, J. and {B{\'e}jar}, V.~J.~S. and {Baroch}, D. and {Bastelberger}, S.~T. and {Chaturvedi}, P. and {Cifuentes}, C. and {Dreizler}, S. and {Jeffers}, S.~V. and {Kopparapu}, R. and {Lafarga}, M. and {L{\'o}pez-Gonz{\'a}lez}, M.~J. and {Mart{\'\i}n-Ruiz}, S. and {Montes}, D. and {Morales}, J.~C. and {Pall{\'e}}, E. and {Pavlov}, A. and {Pedraz}, S. and {Perdelwitz}, V. and {P{\'e}rez-Torres}, M. and {Perger}, M. and {Reffert}, S. and {Rodr{\'\i}guez L{\'o}pez}, C. and {Schlecker}, M. and {Sch{\"o}fer}, P. and {Schweitzer}, A. and {Shan}, Y. and {Shields}, A. and {Stock}, S. and {Wolf}, E. and {Zapatero Osorio}, M.~R. and {Zechmeister}, M.},
        title = "{The CARMENES search for exoplanets around M dwarfs. Wolf 1069 b: Earth-mass planet in the habitable zone of a nearby, very low-mass star}",
      journal = {\aap},
         year = 2023,
        month = feb,
       volume = {670},
          eid = {A84},
        pages = {A84},
          doi = {10.1051/0004-6361/202245322},
archivePrefix = {arXiv},
       eprint = {2301.02477},
 primaryClass = {astro-ph.EP},
       adsurl = {https://ui.adsabs.harvard.edu/abs/2023A&A...670A..84K}
}

@ARTICLE{2008ApJ...676.1262B,
       author = {{Browning}, Matthew K.},
        title = "{Simulations of Dynamo Action in Fully Convective Stars}",
      journal = {\apj},
         year = 2008,
        month = apr,
       volume = {676},
       number = {2},
        pages = {1262-1280},
          doi = {10.1086/527432},
archivePrefix = {arXiv},
       eprint = {0712.1603},
 primaryClass = {astro-ph},
       adsurl = {https://ui.adsabs.harvard.edu/abs/2008ApJ...676.1262B}
}

@ARTICLE{2015ApJ...813L..31Y,
       author = {{Yadav}, Rakesh K. and {Christensen}, Ulrich R. and {Morin}, Julien and {Gastine}, Thomas and {Reiners}, Ansgar and {Poppenhaeger}, Katja and {Wolk}, Scott J.},
        title = "{Explaining the Coexistence of Large-scale and Small-scale Magnetic Fields in Fully Convective Stars}",
      journal = {\apjl},
         year = 2015,
        month = nov,
       volume = {813},
       number = {2},
          eid = {L31},
        pages = {L31},
          doi = {10.1088/2041-8205/813/2/L31},
archivePrefix = {arXiv},
       eprint = {1510.05541},
 primaryClass = {astro-ph.SR},
       adsurl = {https://ui.adsabs.harvard.edu/abs/2015ApJ...813L..31Y}
}

@ARTICLE{2018A&A...610A..19R,
       author = {{Rajpurohit}, A.~S. and {Allard}, F. and {Teixeira}, G.~D.~C. and {Homeier}, D. and {Rajpurohit}, S. and {Mousis}, O.},
        title = "{Photospheric properties and fundamental parameters of M dwarfs}",
      journal = {\aap},
         year = 2018,
        month = feb,
       volume = {610},
          eid = {A19},
        pages = {A19},
          doi = {10.1051/0004-6361/201731507},
archivePrefix = {arXiv},
       eprint = {1708.06211},
 primaryClass = {astro-ph.SR},
       adsurl = {https://ui.adsabs.harvard.edu/abs/2018A&A...610A..19R}
}

@ARTICLE{2021A&A...656A.162M,
       author = {{Marfil}, E. and {Tabernero}, H.~M. and {Montes}, D. and {Caballero}, J.~A. and {L{\'a}zaro}, F.~J. and {Gonz{\'a}lez Hern{\'a}ndez}, J.~I. and {Nagel}, E. and {Passegger}, V.~M. and {Schweitzer}, A. and {Ribas}, I. and {Reiners}, A. and {Quirrenbach}, A. and {Amado}, P.~J. and {Cifuentes}, C. and {Cort{\'e}s-Contreras}, M. and {Dreizler}, S. and {Duque-Arribas}, C. and {Galad{\'\i}-Enr{\'\i}quez}, D. and {Henning}, Th. and {Jeffers}, S.~V. and {Kaminski}, A. and {K{\"u}rster}, M. and {Lafarga}, M. and {L{\'o}pez-Gallifa}, {\'A}. and {Morales}, J.~C. and {Shan}, Y. and {Zechmeister}, M.},
        title = "{The CARMENES search for exoplanets around M dwarfs. Stellar atmospheric parameters of target stars with SteParSyn}",
      journal = {\aap},
         year = 2021,
        month = dec,
       volume = {656},
          eid = {A162},
        pages = {A162},
          doi = {10.1051/0004-6361/202141980},
archivePrefix = {arXiv},
       eprint = {2110.07329},
 primaryClass = {astro-ph.SR},
       adsurl = {https://ui.adsabs.harvard.edu/abs/2021A&A...656A.162M}
}

@ARTICLE{2021A&A...649A.103O,
       author = {{Olander}, T. and {Heiter}, U. and {Kochukhov}, O.},
        title = "{Comparative high-resolution spectroscopy of M dwarfs: Exploring non-LTE effects}",
      journal = {\aap},
         year = 2021,
        month = may,
       volume = {649},
          eid = {A103},
        pages = {A103},
          doi = {10.1051/0004-6361/202039747},
archivePrefix = {arXiv},
       eprint = {2102.08836},
 primaryClass = {astro-ph.SR},
       adsurl = {https://ui.adsabs.harvard.edu/abs/2021A&A...649A.103O}
}

@ARTICLE{2021A&ARv..29....1K,
       author = {{Kochukhov}, Oleg},
        title = "{Magnetic fields of M dwarfs}",
      journal = {\aapr},
         year = 2021,
        month = dec,
       volume = {29},
       number = {1},
          eid = {1},
        pages = {1},
          doi = {10.1007/s00159-020-00130-3},
archivePrefix = {arXiv},
       eprint = {2011.01781},
 primaryClass = {astro-ph.SR},
       adsurl = {https://ui.adsabs.harvard.edu/abs/2021A&ARv..29....1K}
}

@ARTICLE{2019A&A...631A.100C,
       author = {{Chiavassa}, A. and {Brogi}, M.},
        title = "{Planet and star synergy at high-spectral resolution. A rationale for the characterization of exoplanet atmospheres. I. The infrared}",
      journal = {\aap},
         year = 2019,
        month = nov,
       volume = {631},
          eid = {A100},
        pages = {A100},
          doi = {10.1051/0004-6361/201936566},
archivePrefix = {arXiv},
       eprint = {1909.11807},
 primaryClass = {astro-ph.EP},
       adsurl = {https://ui.adsabs.harvard.edu/abs/2019A&A...631A.100C}
}

@ARTICLE{2024MNRAS.528.3509V,
       author = {{Vaughan}, Sophia R. and {Birkby}, Jayne L. and {Thatte}, Niranjan and {Carlotti}, Alexis and {Houll{\'e}}, Mathis and {Pereira-Santaella}, Miguel and {Clarke}, Fraser and {Vigan}, Arthur and {Lin}, Zifan and {Kaltenegger}, Lisa},
        title = "{Behind the mask: can HARMONI@ELT detect biosignatures in the reflected light of Proxima b?}",
      journal = {\mnras},
         year = 2024,
        month = feb,
       volume = {528},
       number = {2},
        pages = {3509-3522},
          doi = {10.1093/mnras/stae242},
archivePrefix = {arXiv},
       eprint = {2401.09589},
 primaryClass = {astro-ph.EP},
       adsurl = {https://ui.adsabs.harvard.edu/abs/2024MNRAS.528.3509V}
}

@ARTICLE{2025ExA....59...29P,
       author = {{Palle}, Enric and {Biazzo}, Katia and {Bolmont}, Emeline and {Molli{\`e}re}, Paul and {Poppenhaeger}, Katja and {Birkby}, Jayne and {Brogi}, Matteo and {Chauvin}, Gael and {Chiavassa}, Andrea and {Hoeijmakers}, Jens and {Lellouch}, Emmanuel and {Lovis}, Christophe and {Maiolino}, Roberto and {Nortmann}, Lisa and {Parviainen}, Hannu and {Pino}, Lorenzo and {Turbet}, Martin and {Weder}, Jesse and {Albrecht}, Simon and {Antoniucci}, Simone and {Barros}, Susana C. and {Beaudoin}, Andre and {Benneke}, Bjorn and {Boisse}, Isabelle and {Bonomo}, Aldo S. and {Borsa}, Francesco and {Brandeker}, Alexis and {Brandner}, Wolfgang and {Buchhave}, Lars A. and {Cheffot}, Anne-Laure and {Deborde}, Robin and {Debras}, Florian and {Doyon}, Rene and {Di Marcantonio}, Paolo and {Giacobbe}, Paolo and {Gonz{\'a}lez Hern{\'a}ndez}, Jonay I. and {Helled}, Ravit and {Kreidberg}, Laura and {Machado}, Pedro and {Maldonado}, Jesus and {Marconi}, Alessandro and {Martins}, B.~L. Canto and {Miceli}, Adriano and {Mordasini}, Christoph and {N'Diaye}, Mamadou and {Niedzielski}, Andrzej and {Nisini}, Brunella and {Origlia}, Livia and {Peroux}, Celine and {Pietrow}, Alexander G.~M. and {Pinna}, Enrico and {Rauscher}, Emily and {Reffert}, Sabine and {Rodr{\'\i}guez-L{\'o}pez}, Cristina and {Rousselot}, Philippe and {Sanna}, Nicoletta and {Santos}, Nuno C. and {Simonnin}, Adrien and {Su{\'a}rez Mascare{\~n}o}, Alejandro and {Zanutta}, Alessio and {Zapatero-Osorio}, Maria Rosa and {Zechmeister}, Mathias},
        title = "{Ground-breaking exoplanet science with the ANDES spectrograph at the ELT}",
      journal = {Experimental Astronomy},
         year = 2025,
        month = jun,
       volume = {59},
       number = {3},
          eid = {29},
        pages = {29},
          doi = {10.1007/s10686-025-10000-4},
archivePrefix = {arXiv},
       eprint = {2311.17075},
 primaryClass = {astro-ph.IM},
       adsurl = {https://ui.adsabs.harvard.edu/abs/2025ExA....59...29P}
}

@ARTICLE{2011A&A...535A..22C,
       author = {{Chiavassa}, A. and {Freytag}, B. and {Masseron}, T. and {Plez}, B.},
        title = "{Radiative hydrodynamics simulations of red supergiant stars. IV. Gray versus non-gray opacities}",
      journal = {\aap},
         year = 2011,
        month = nov,
       volume = {535},
          eid = {A22},
        pages = {A22},
          doi = {10.1051/0004-6361/201117463},
archivePrefix = {arXiv},
       eprint = {1109.3619},
 primaryClass = {astro-ph.SR},
       adsurl = {https://ui.adsabs.harvard.edu/abs/2011A&A...535A..22C}
}

@ARTICLE{2009MmSAI..80..731S,
       author = {{Steffen}, M. and {Ludwig}, H.-G. and {Caffau}, E.},
        title = "{Micro- and macroturbulence derived from 3D hydrodynamical stellar atmospheres .}",
      journal = {\memsai},
         year = 2009,
        month = jan,
       volume = {80},
        pages = {731},
          doi = {10.48550/arXiv.0909.2831},
archivePrefix = {arXiv},
       eprint = {0909.2831},
 primaryClass = {astro-ph.SR},
       adsurl = {https://ui.adsabs.harvard.edu/abs/2009MmSAI..80..731S}
}

@ARTICLE{2015ApJ...804...64M,
       author = {{Mann}, Andrew W. and {Feiden}, Gregory A. and {Gaidos}, Eric and {Boyajian}, Tabetha and {von Braun}, Kaspar},
        title = "{How to Constrain Your M Dwarf: Measuring Effective Temperature, Bolometric Luminosity, Mass, and Radius}",
      journal = {\apj},
         year = 2015,
        month = may,
       volume = {804},
       number = {1},
          eid = {64},
        pages = {64},
          doi = {10.1088/0004-637X/804/1/64},
archivePrefix = {arXiv},
       eprint = {1501.01635},
 primaryClass = {astro-ph.SR},
       adsurl = {https://ui.adsabs.harvard.edu/abs/2015ApJ...804...64M}
}

@ARTICLE{2021Msngr.182...38K,
       author = {{Kasper}, M. and {Cerpa Urra}, N. and {Pathak}, P. and {Bonse}, M. and {Nousiainen}, J. and {Engler}, B. and {Heritier}, C.~T. and {Kammerer}, J. and {Leveratto}, S. and {Rajani}, C. and {Bristow}, P. and {Le Louarn}, M. and {Madec}, P.-Y. and {Str{\"o}bele}, S. and {Verinaud}, C. and {Glauser}, A. and {Quanz}, S.~P. and {Helin}, T. and {Keller}, C. and {Snik}, F. and {Boccaletti}, A. and {Chauvin}, G. and {Mouillet}, D. and {Kulcs{\'a}r}, C. and {Raynaud}, H.-F.},
        title = "{PCS {\textemdash} A Roadmap for Exoearth Imaging with the ELT}",
      journal = {The Messenger},
         year = 2021,
        month = mar,
       volume = {182},
        pages = {38-43},
          doi = {10.18727/0722-6691/5221},
archivePrefix = {arXiv},
       eprint = {2103.11196},
 primaryClass = {astro-ph.IM},
       adsurl = {https://ui.adsabs.harvard.edu/abs/2021Msngr.182...38K}
}

@ARTICLE{2021Msngr.182...22B,
       author = {{Brandl}, B. and {Bettonvil}, F. and {van Boekel}, R. and {Glauser}, A. and {Quanz}, S. and {Absil}, O. and {Amorim}, A. and {Feldt}, M. and {Glasse}, A. and {G{\"u}del}, M. and {Ho}, P. and {Labadie}, L. and {Meyer}, M. and {Pantin}, E. and {van Winckel}, H. and {METIS Consortium}},
        title = "{METIS: The Mid-infrared ELT Imager and Spectrograph}",
      journal = {The Messenger},
         year = 2021,
        month = mar,
       volume = {182},
        pages = {22-26},
          doi = {10.18727/0722-6691/5218},
archivePrefix = {arXiv},
       eprint = {2103.11208},
 primaryClass = {astro-ph.IM},
       adsurl = {https://ui.adsabs.harvard.edu/abs/2021Msngr.182...22B}
}

@ARTICLE{2025ARA&A..63...83S,
       author = {{Snellen}, Ignas A.~G.},
        title = "{Exoplanet Atmospheres at High Spectral Resolution}",
      journal = {\araa},
         year = 2025,
        month = aug,
       volume = {63},
       number = {1},
        pages = {83-125},
          doi = {10.1146/annurev-astro-052622-031342},
archivePrefix = {arXiv},
       eprint = {2505.08926},
 primaryClass = {astro-ph.EP},
       adsurl = {https://ui.adsabs.harvard.edu/abs/2025ARA&A..63...83S}
}

@ARTICLE{2017A&A...599A..16L,
       author = {{Lovis}, C. and {Snellen}, I. and {Mouillet}, D. and {Pepe}, F. and {Wildi}, F. and {Astudillo-Defru}, N. and {Beuzit}, J.-L. and {Bonfils}, X. and {Cheetham}, A. and {Conod}, U. and {Delfosse}, X. and {Ehrenreich}, D. and {Figueira}, P. and {Forveille}, T. and {Martins}, J.~H.~C. and {Quanz}, S.~P. and {Santos}, N.~C. and {Schmid}, H.-M. and {S{\'e}gransan}, D. and {Udry}, S.},
        title = "{Atmospheric characterization of Proxima b by coupling the SPHERE high-contrast imager to the ESPRESSO spectrograph}",
      journal = {\aap},
         year = 2017,
        month = mar,
       volume = {599},
          eid = {A16},
        pages = {A16},
          doi = {10.1051/0004-6361/201629682},
archivePrefix = {arXiv},
       eprint = {1609.03082},
 primaryClass = {astro-ph.EP},
       adsurl = {https://ui.adsabs.harvard.edu/abs/2017A&A...599A..16L}
}

@ARTICLE{2015A&A...576A..59S,
       author = {{Snellen}, I. and {de Kok}, R. and {Birkby}, J.~L. and {Brandl}, B. and {Brogi}, M. and {Keller}, C. and {Kenworthy}, M. and {Schwarz}, H. and {Stuik}, R.},
        title = "{Combining high-dispersion spectroscopy with high contrast imaging: Probing rocky planets around our nearest neighbors}",
      journal = {\aap},
         year = 2015,
        month = apr,
       volume = {576},
          eid = {A59},
        pages = {A59},
          doi = {10.1051/0004-6361/201425018},
archivePrefix = {arXiv},
       eprint = {1503.01136},
 primaryClass = {astro-ph.EP},
       adsurl = {https://ui.adsabs.harvard.edu/abs/2015A&A...576A..59S}
}

@ARTICLE{Kosiarek2019,
       author = {{Kosiarek}, Molly R. and {Crossfield}, Ian J.~M. and {Hardegree-Ullman}, Kevin K. and {Livingston}, John H. and {Benneke}, Bj{\"o}rn and {Henry}, Gregory W. and {Howard}, Ward S. and {Berardo}, David and {Blunt}, Sarah and {Fulton}, Benjamin J. and {Hirsch}, Lea A. and {Howard}, Andrew W. and {Isaacson}, Howard and {Petigura}, Erik A. and {Sinukoff}, Evan and {Weiss}, Lauren and {Bonfils}, X. and {Dressing}, Courtney D. and {Knutson}, Heather A. and {Schlieder}, Joshua E. and {Werner}, Michael and {Gorjian}, Varoujan and {Krick}, Jessica and {Morales}, Farisa Y. and {Astudillo-Defru}, Nicola and {Almenara}, J.-M. and {Delfosse}, X. and {Forveille}, T. and {Lovis}, C. and {Mayor}, M. and {Murgas}, F. and {Pepe}, F. and {Santos}, N.~C. and {Udry}, S. and {Corbett}, H.~T. and {Fors}, Octavi and {Law}, Nicholas M. and {Ratzloff}, Jeffrey K. and {del Ser}, Daniel},
        title = "{Bright Opportunities for Atmospheric Characterization of Small Planets: Masses and Radii of K2-3 b, c, and d and GJ3470 b from Radial Velocity Measurements and Spitzer Transits}",
      journal = {\aj},
         year = 2019,
        month = mar,
       volume = {157},
       number = {3},
          eid = {97},
        pages = {97},
          doi = {10.3847/1538-3881/aaf79c},
archivePrefix = {arXiv},
       eprint = {1812.08241},
 primaryClass = {astro-ph.EP},
       adsurl = {https://ui.adsabs.harvard.edu/abs/2019AJ....157...97K}
}

@ARTICLE{Flowers2019,
       author = {{Flowers}, Erin and {Brogi}, Matteo and {Rauscher}, Emily and {Kempton}, Eliza M.-R. and {Chiavassa}, Andrea},
        title = "{The High-resolution Transmission Spectrum of HD 189733b Interpreted with Atmospheric Doppler Shifts from Three-dimensional General Circulation Models}",
      journal = {\aj},
         year = 2019,
        month = may,
       volume = {157},
       number = {5},
          eid = {209},
        pages = {209},
          doi = {10.3847/1538-3881/ab164c},
archivePrefix = {arXiv},
       eprint = {1810.06099},
 primaryClass = {astro-ph.EP},
       adsurl = {https://ui.adsabs.harvard.edu/abs/2019AJ....157..209F}
}

@ARTICLE{Welbanks2024,
       author = {{Welbanks}, Luis and {Bell}, Taylor J. and {Beatty}, Thomas G. and {Line}, Michael R. and {Ohno}, Kazumasa and {Fortney}, Jonathan J. and {Schlawin}, Everett and {Greene}, Thomas P. and {Rauscher}, Emily and {McGill}, Peter and {Murphy}, Matthew and {Parmentier}, Vivien and {Tang}, Yao and {Edelman}, Isaac and {Mukherjee}, Sagnick and {Wiser}, Lindsey S. and {Lagage}, Pierre-Olivier and {Dyrek}, Achr{\`e}ne and {Triantafillides}, Anastasia},
        title = "{A high internal heat flux and large core in a warm Neptune exoplanet}",
      journal = {\nat},
         year = 2024,
        month = jun,
       volume = {630},
       number = {8018},
        pages = {836-840},
          doi = {10.1038/s41586-024-07514-w},
archivePrefix = {arXiv},
       eprint = {2405.11018},
 primaryClass = {astro-ph.EP},
       adsurl = {https://ui.adsabs.harvard.edu/abs/2024Natur.630..836W}
}

\begin{appendix}

\section{Inclusion of the full stellar-orbital spin angle in the CLV+RME model}\label{sec:appendix_clv-rme_model}

To find a simple expression for the line of sight (LOS) velocity of a rotating star, we review the coordinate system of our transit model, which is centered with the star, with ($x$,$y$) coordinates on the plane of the sky (pointing right and up, respectively) and the $z$ axis pointing along the LOS towards the observer.

For a completely aligned system ($\lambda=0^\circ$, $i_\star=90^\circ$), the angular velocity vector is parallel to the $y$ axis, and thus it has coordinates:
\begin{equation}
 \vec{\omega}_0 = [0,v_{eq}/R_\star,0] = [0,v_{eq},0],   
\end{equation}
where we used our choice of $R_\star=1$. We apply two rotation matrices to $\vec{\omega}_0$, one clockwise around the $z$ axis to account for the obliquity $\lambda$, and the second counter-clockwise around the $x$ axis to account for the stellar inclination $i_\star$. Doing so, we obtain a rotated angular velocity vector of coordinates:
\begin{equation}
   \vec{\omega} \equiv v_{eq}[\sin\lambda\sin i_\star, \cos\lambda\sin i_\star,\cos i_\star] 
\end{equation}
If each point on the rotating stellar surface has position vector $\vec{r} = [x,y,z]$, its velocity vector is given by the cross product $\vec{v} = \vec{\omega} \times \vec{r}$. Thus, the LOS velocity $\vec{v}_{LOS}$ is the $z$ component of the cross product, that is $\vec{v}_{LOS} = (\omega_xy-\omega_yx)\,\vec{u}_z$. Substituting the expression for $\vec{\omega}$:
\begin{equation}
    v_{LOS}(x,y) = v_{eq}\sin{i_\star}(y\,\sin\lambda - x\,\cos\lambda),
\end{equation}
where $(x,y)$ are the grid coordinates on the plane of the sky, expressed in units of stellar radius as in our model. The radial velocity is obtained by changing the sign of $v_{LOS}$, because by convention it is positive along the $-z$ axis (i.e. when the object moves away from the observer).

In practical application, from stellar spectroscopy the quantity that is commonly measured is not the modulus of the equatorial velocity $v_{eq}$ but rather its projected value $v\sin i \equiv v_{eq} \sin i_\star$ in our equations above. Hence, the final radial velocity is given by:
\begin{equation}\label{eq:v_LOS}
    RV(x,y) = -v\sin i\,(y\,\sin\lambda - x\,\cos\lambda),
\end{equation}
which is used to Doppler shift the spectrum corresponding to each gridded pixel of the stellar surface.

\section{SVD analysis on separate stellar and telluric effects}

\begin{figure*}
    \centering
    \includegraphics[width=0.8\hsize]{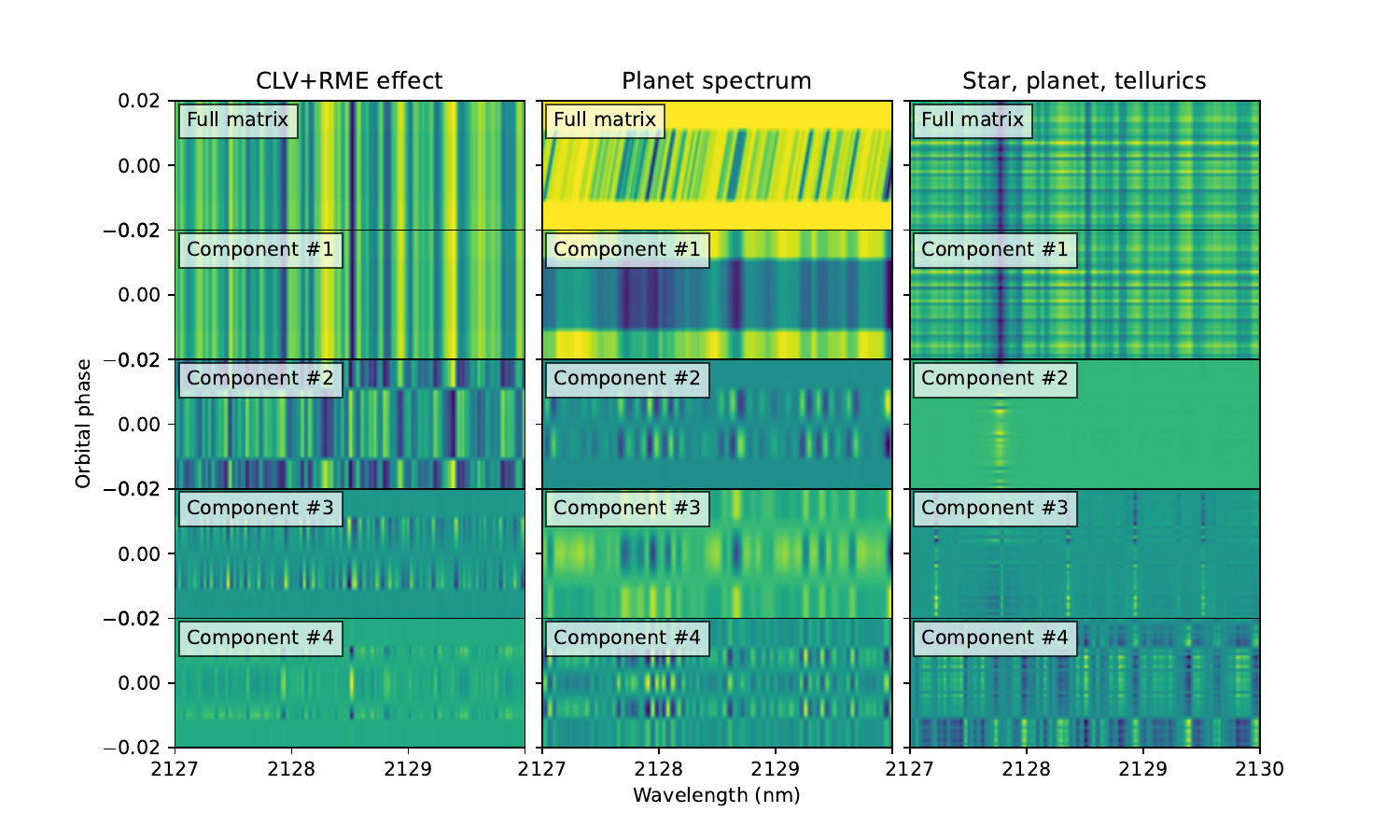}
    \caption{Right panels: same as the noiseless case in Fig. \ref{fig:pca_components}, except that we also show the reconstructed SVD components when simulating the stellar effects only (left panel) and the exoplanet transmission spectrum only (middle panel). See Sect.~\ref{sec:PCA} for details.}
    \label{fig:svd_separate_effects}
\end{figure*}

In Fig. \ref{fig:svd_separate_effects} we show how different astrophysical effects correspond to different SVD components found when decomposing the (wavelength, time) matrices. It shows that it is hard to predict the behaviour and thus the efficacy of SVD in detrending such effects without a dedicated simulation for each dataset, as argued in Sect.~\ref{sec:clv+rme_model}.

\section{Recovery of the transmission spectrum of GJ 3470b post-SVD}\label{sec:appendix_detection_tests}

In this section, we study whether residual uncorrected RME+CLVs even after the application of SVD as explained in Sect.~\ref{sec:clv+rme_model} can hamper or otherwise bias the detection of the transmission spectrum of GJ 3470\,b, given that planet lines and residual stellar effects have similar amplitudes.

\begin{figure*}
    \centering
    \includegraphics[width=1.1\hsize,trim=0cm 1cm 0cm 1cm]{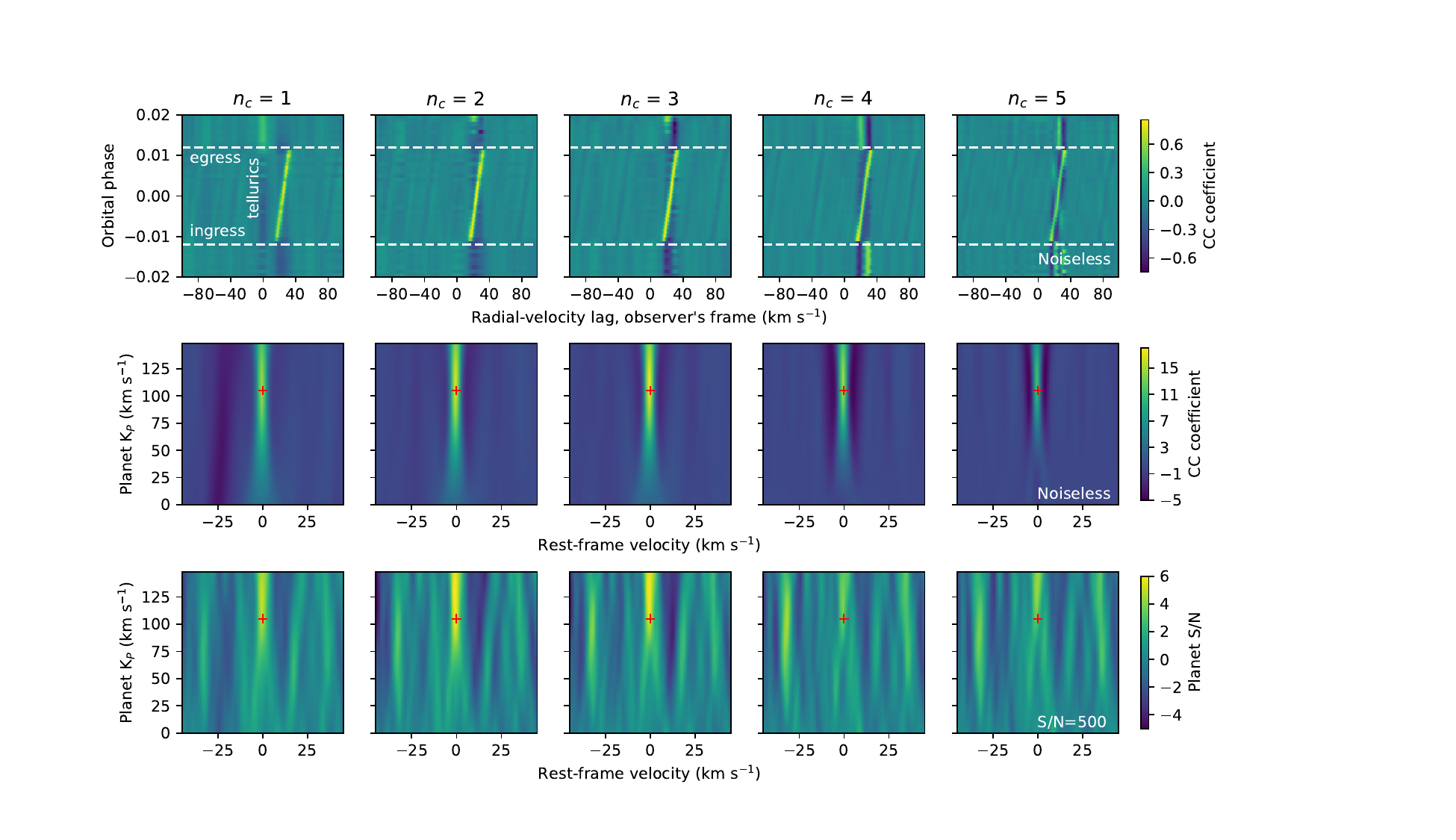}
    \caption{Top panel: trail matrices obtained by correlating the noiseless SVD residuals in Sect.~\ref{sec:PCA} with the transmission spectrum of GJ\,3470\,b, as a function of planet orbital phase and radial velocity lag in the observer's frame. Middle panel: total cross correlation coefficients obtained by shifting the noiseless trail matrices of the top row into the planet rest frame and co-adding in phase. Signals are shown as a function of rest frame velocity and orbital radial velocity semi-amplitude ($K_P$). Bottom panel: as in the middle panel, but for a simulation with S/N=500 per sampling element. From left to right, the figure shows increasing number of SVD components ($n_c$) removed.}
    \label{fig:post-svd_detections}
\end{figure*}

To offer a visualization of the SVD performances, we compute \textit{trail matrices} in the noiseless case. These matrices contain the cross correlation coefficients calculated between the SVD residuals and the model of the exoplanet transmission spectrum. In the top panel of Fig. \ref{fig:post-svd_detections}, it is possible to see that even for just 1 SVD component ($n_c=1)$ removed, apart from a spurious telluric correlation residual at zero RV lag, the planet's trail (excess correlation) is correctly recovered. In the middle panel, we indeed demonstrate that the maximum cross correlation matches the simulated parameters of GJ 3470\,b in the ($V_{rest}$,$K_P$) maps. 

Interestingly, for increasing $n_c$, we observe a progressive smearing of the exoplanet signal in phase, which is documented in the literature \citep{Dash2024}. Such smearing transfers a portion of the exoplanet signal from in-transit to out-of-transit phases. This is a potentially tricky effect, which has to be accounted for by reprocessing each of the tested models as closely as possible to the data \citep{Gibson2022}, and complicates the proposed substitution of the $K_P$ parameter with transit duration \citep{Cheverall2024}.

Lastly, we ran a recovery test on noisy data (Fig.~\ref{fig:post-svd_detections}, bottom panel). Due to the limited spectral range of these simulations (2125--2145 nm) we can only obtain a detection by increasing the S/N per spectral pixel to 500. For comparison, the S/N per pixel of the GIANO observations presented in Sect.~\ref{sec:model-data_comparison} is about 35, but GIANO covers the whole 0.95--2.45 $\mu$m spectral range almost continuously. Even in the noisy simulations, there doesn't seem to be any major distortion or systematic effect from the stellar RME+CLVs independently of the number of components. The measured planet S/N (calculated by dividing the peak cross correlation through the standard deviation of the correlation values away from the peak) varies a bit irregularly with $n_c$, which might point to some residual stellar effects having a marginal quantitative effect. We also note that the noise seed is fixed here, so every time the simulation is repeated the underlying noise structure is unaltered. Every difference is thus entirely due to SVD.

\end{appendix}

\end{document}